\documentclass[fleqn,usenatbib]{mnras}

\usepackage{newtxtext}%,newtxmath}
\usepackage[T1]{fontenc}
\DeclareRobustCommand{\VAN}[3]{#2}
\let\VANthebibliography\thebibliography
\def\thebibliography{\DeclareRobustCommand{\VAN}[3]{##3}\VANthebibliography}

\usepackage{graphicx}	% Including figure files
\usepackage{amsmath}	% Advanced maths commands
\usepackage{amssymb}	% Extra maths symbols
\usepackage{mathptmx}
\usepackage{txfonts} %this has to load after amsmath and amssymb because latex is dumb
\usepackage{subcaption} %multi-panel figures
\usepackage{hyperref}
\usepackage{pdflscape}
\usepackage{multirow} %multiple rows labeled in tables
\usepackage{soul} %allows /st{}

\usepackage{color}
\setstcolor{magenta}
\definecolor{darkgreen}{rgb}{0,0.5,0}
\definecolor{mag}{rgb}{0.79,0.08,0.48}
\definecolor{darkblue}{rgb}{0,0.2,0.8}
\definecolor{cyan}{rgb}{0,0.8,0.8}

\title[Ellipse+shear modeling assumption on $H_0$]{Galaxy-lens determination of $H_0$: the effect of the ellipse+shear modeling assumption}
\author[Gomer and Williams]{
Matthew R. Gomer,
Liliya L. R. Williams
\\
School of Physics and Astronomy, University of Minnesota, 116 Church Street SE, Minneapolis MN, 55455, USA\\
}

\date{Accepted XXX. Received YYY; in original form ZZZ}

\pubyear{2021}

\begin{document}
\label{firstpage}
\pagerange{\pageref{firstpage}--\pageref{lastpage}}
\maketitle

\begin{abstract}
    Galaxy lenses are frequently modeled as an elliptical mass distribution with external shear and isothermal spheres to account for secondary and line-of-sight galaxies. There is statistical evidence that some fraction of observed quads are inconsistent with these assumptions, and require a dipole-like contribution to the mass with respect to the light. Simplifying assumptions about the shape of mass distributions can lead to the incorrect recovery of parameters such as $H_0$. We create several tests of synthetic quad populations with different deviations from an elliptical shape, then fit them with an ellipse+shear model, and measure the recovered values of $H_0$.  Kinematic constraints are not included. We perform two types of fittings- one with a single point source and one with an array of sources emulating an extended source. We carry out two model-free comparisons between our mock quads and the observed population. One result of these comparisons is a statistical inconsistency not yet mentioned in the literature: the image distance ratios with respect to the lens center of observed quads appear to  span a much wider range than those of synthetic or simulated quads. Bearing this discrepancy in mind, our mock populations can result in biases on $H_0$ $\sim10\%$.
\end{abstract}


% Abstract of the paper

% Select between one and six entries from the list of approved keywords.
% Don't make up new ones.

% \begin{document}

% \title[Ellipse+shear modeling assumption on $H_0$]{Galaxy-lens determination of $H_0$: the effect of the ellipse+shear modeling assumption}
% \title{Galaxy-lens determination of $H_0$: the effect of the ellipse+shear modeling assumption}

%% %simple case: 2 authors, same institution
%% \author{A. Uthor}
%% \author{and A. Nother Author}
%% \affiliation{Institution,\\Address, Country}

% more complex case: 4 authors, 3 institutions, 2 footnotes
% \author[a,1]{M. Gomer,\note{Corresponding author.}}
% \author[a]{L. L. R. Williams,}

% The "\note" macro will give a warning: "Ignoring empty anchor..."
% you can safely ignore it.

% \affiliation[a]{University of Minnesota\\ 116 Church Street SE, Minneapolis MN, 55455, USA}

% e-mail addresses: one for each author, in the same order as the authors
% \emailAdd{gomer011@umn.edu}
% \emailAdd{llrw@umn.edu}

% \begin{document}
% \maketitle
% \flushbottom

\section{Introduction}
The two major competing methods to measure $H_0$, through temperature anisotropies of the CMB and standard candle distance determinations, currently disagree at the $4.4\sigma$ level \citep{Planck18,Riess19}. To diagnose or potentially resolve this tension, the gold standard is to measure $H_0$ to 1\% precision. One method which may be competitive for this goal is to use time delays from strong gravitational lensing as a direct measure of distance. The most precise constraint from this method to date comes from the H0LiCOW ($H_0$ Lenses in COSMOGRAIL's Wellspring) program \citep{HC13}, who recently used a combined analysis of six lens systems to find $H_0=73.3 \substack{+1.7 \\ -1.8}$ km s$^{-1}$Mpc$^{-1}$ (2.4\% uncertainty), in agreement with the \citet{Riess19} standard candle value.

The lensing method works by measuring the difference in arrival time between two or more images, which arises due to the paths having different lengths and passing through different gravitational potentials. This determination provides a direct measure of a combination of distances, and therefore is directly related to $H_0$: 
$D_{\Delta t}=(1+z_d)\frac{D_d D_s}{D_{ds}} \propto \frac{1}{H_0}$ \citep{Refsdal64, Schechter97}. 
The accuracy of this determination can only be as good as the measurement of the time delays and the accuracy of the lens model. As such, the H0LiCOW group has gone to great efforts to precisely model each of their lens systems. Time delays are measured from the COSMOGRAIL program, a long-term monitoring program of multiply imaged quasars, which has measured time delays to within 1-3\% \citep{Courbin04,Bonvin16}. The main lens is modeled as an ellipse+shear, either as a power law profile or a composite profile with a baryonic component and a dark matter (DM) NFW component \citep{NFW96}, and a Bayesian inference is used to choose the best model \citep{HC13}. Line-of-sight and neighboring galaxies are included in the modeling process. Large-scale smooth line-of-sight structure is accounted for through a statistical comparison with control surveys and simulations. Stellar kinematic information of the lens is used to constrain the mass of the system, breaking the Mass Sheet Degeneracy (MSD). 

\subsection{Lensing degeneracies}
Despite this enormous effort, there is still room for uncertainty. Gravitational lensing is plagued by many degeneracies, where the same observables can be reproduced by a family of lenses. The most famous is the aforementioned Mass Sheet Degeneracy (MSD), \citep{Falco85, Saha00}, where scaling of the convergence ($\Sigma/\Sigma_{crit}$) by a factor of $\lambda$ and adding a uniform convergence of $(1-\lambda)$ does not affect the image positions or the
relative fluxes.

\begin{equation}
\kappa_{\lambda}(\vec{x})=\lambda\kappa(\vec{x})+(1-\lambda)
\end{equation}

However, the relative time delays are affected by a factor of $\lambda$, which in turn means the recovered value of $H_0$ will be biased by a factor of $\lambda$. In principle, any value of $\lambda$ is equally well supported by the lensing data, but one particular value is artificially selected in the modeling process. If the effect is similar for many systems, this could impart a bias on the recovered value of $H_0$ \citep{Schneider13,Xu16}. Phrased another way, the choice of lens model may select the value of $\lambda$ which causes the recovered lens to most closely match that choice model, whether or not that corresponds to the true mass distribution. No model mass distribution will ever perfectly match the true mass distribution of a particular lens, so this effect is always present to some degree. In this way, since the mass distributions of real galaxies will always be more intricate than our lens models, the simplifying assumptions made in the construction of those models may introduce systematic effects in the recovery of parameters like $H_0$.

The MSD in particular is one of the more-studied lensing degeneracies. Recent work by \citet{TDCOSMO4,TDCOSMO5} explicitly folds uncertainties related to the MSD into the analysis of H0LiCOW lenses, resulting in a constraint on $H_0$ of $\sim8\%$. \citet{Xu16} and \citet{Tagore18} extracted halos from the Illustris and EAGLE simulations and examined their lens profiles in a statistical way. The studies envisioned fitting each mock lens profile as a power law and calculating the $\lambda$ necessary to transform each profile into a power-law shape near the image radius. Assuming this distribution of $\lambda$ values would be equivalent to the bias on $H_0$, the authors make statistical determinations as to the bias and spread of $H_0$ recovery for these lenses. \citet{Gomer19} instead explicitly fit mock quads from two-component analytical profiles using a power-law model and found that the bias and spread on $H_0$ was not the same as the $\lambda$ values expected from the above rationale, perhaps casting doubt on the applicability of the statistical distributions of $\lambda$ calculated by \citet{Xu16} and \citet{Tagore18}.

As mentioned above, the H0LiCOW project includes stellar kinematic information to break the MSD. The principle is that the velocity dispersion is measured at distances from the galaxy center where stars dominate, which provides an absolute measure of mass at those radii. Forcing the lens model to match this constraint restricts the freedom of $\lambda$, and therefore $H_0$. \citet{Gomer19} discovered interesting results with regard to this practice. In the fitting procedure, they forced the slope to take on the actual value of the lens profile slope (near the image radius) which serves to emulate the effect of stellar kinematics constraints. Slope constraints and stellar kinematic constraints are similar because in both cases external information about the mass distribution near the image radius is used to inform the fitting process. Strangely, \citet{Gomer19} found that this could introduce significant bias in the recovery of $H_0$. When the value of slope corresponding to that of the actual mass distribution is provided, the MSD is broken, but it is done so incorrectly, so as to introduce bias on $H_0$. Lensing degeneracies continue to surprise us as they manifest in unexpected ways. Simplifying assumptions about the profile shape have caused the modeling process to recover the wrong value of $H_0$, even when informed with external information which should have improved the recovery. This may have consequences for the H0LiCOW determination of $H_0$, or any other determination which uses stellar kinematic information to break the MSD.

Unfortunately the problem is not limited to the MSD. For one, the MSD is actually a special case of the more general source-position transformation (SPT, \citet{Schneider14}), which takes any single source which produces multiple images and describes the possible mappings to reproduce those images using a different source position. With more flexibility than the MST (Mass Sheet Transformation), the image positions and relative magnifications are reproduced exactly in the axisymmetric case. Though the observable quantities are not perfectly reproduced in the general case, they are very nearly matched (within the errors of observations) with realistic ellipticity values. Like the MST, the time delays (and therefore $H_0$) are affected, although unlike the MST they do not scale evenly with the source position, making the effects more complicated to parse. 

\begin{figure}
\begin{tabular}[c]{cc}
\centering
 \begin{subfigure}[c]{.47\linewidth}
   \centering
   \includegraphics[width=\linewidth]{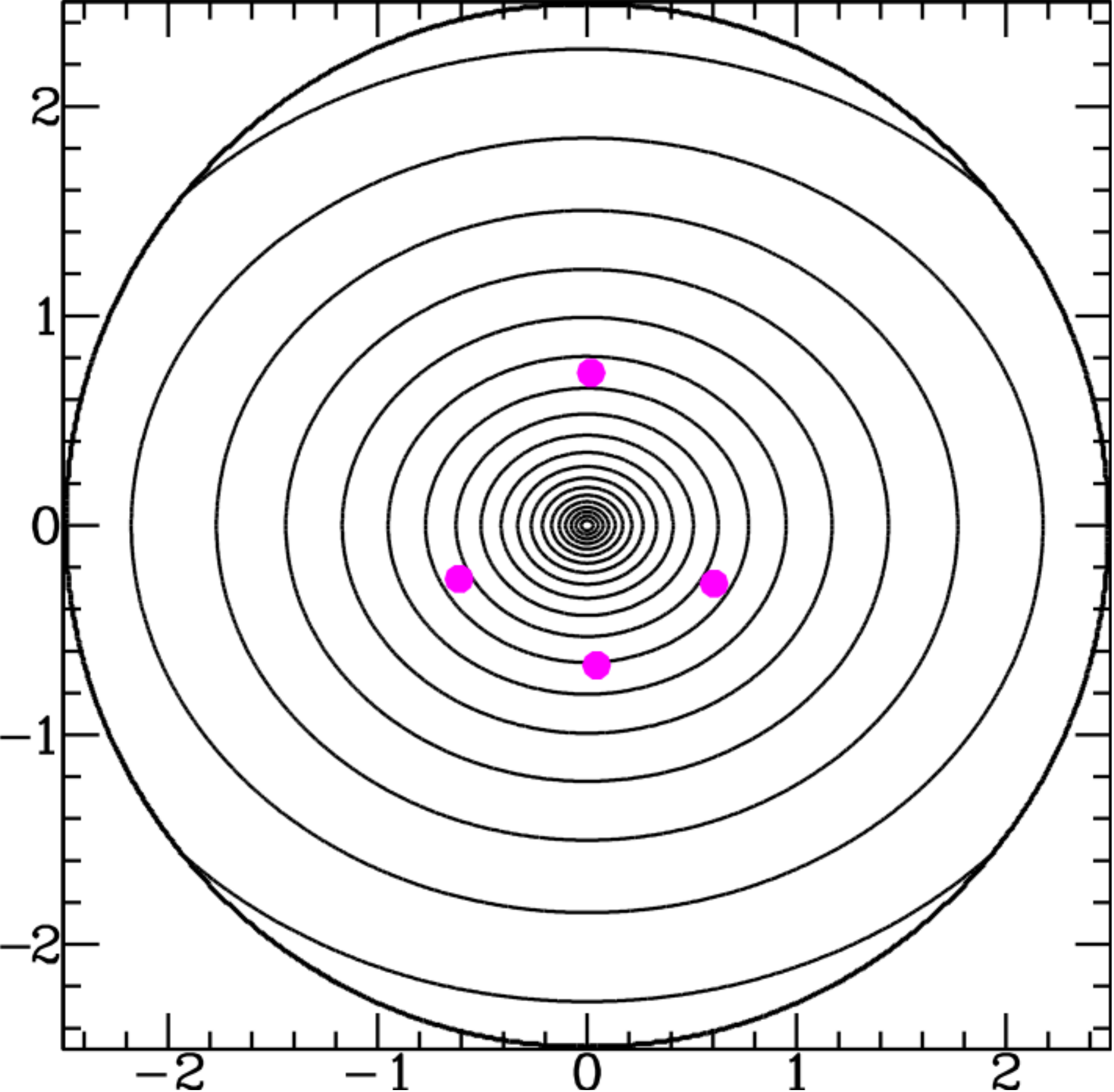}
   \label{fig:sfigmono1}
 \end{subfigure}
 \begin{subfigure}[c]{.47\linewidth}
   \centering
   \includegraphics[width=\linewidth]{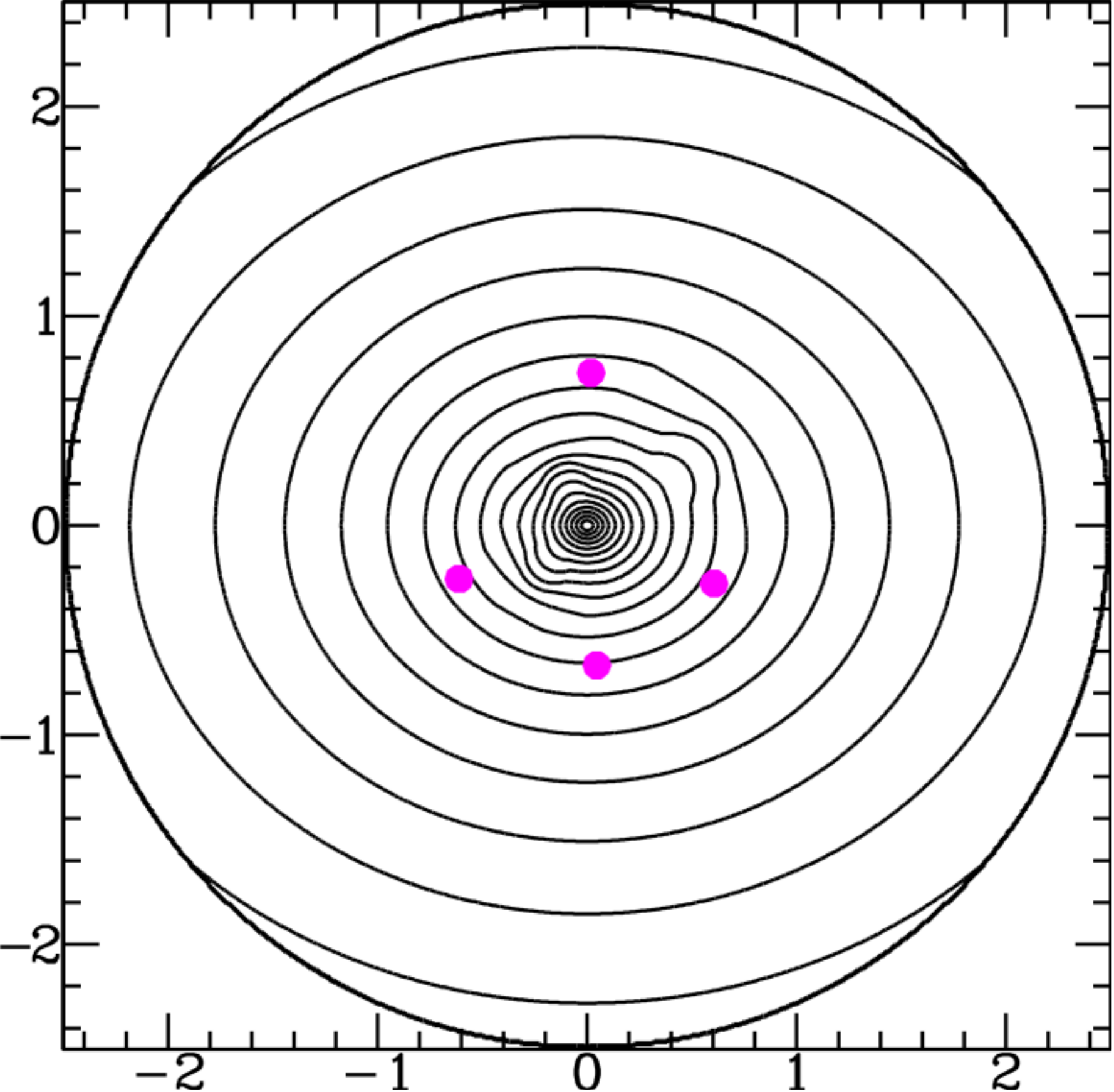}
   \label{fig:sfigmono2}
 \end{subfigure}
\end{tabular}
 \caption{
    An example of the monopole degeneracy. On the left is a simple elliptical mass distribution with the four images from an example quad depicted in magenta (scale in arcsec). On the right is the distribution after three monopoles have been applied. The quad images are in the same location with the same time delays because of the monopole degeneracy. The structure of the mass distribution could be quite complex, but this particular quad would not reveal it. If the true mass distribution were akin the that on the right, models would only ever recover the mass distribution on the left, since they assume a perfectly elliptical lens mass.
    }
\label{fig:monoexample}
\end{figure}

Another known degeneracy is the monopole degeneracy, where any circular region of the 2D mass distribution which does not contain an image can be altered by simultaneously adding and subtracting convergence in a circularly symmetric way (i.e. can be described with a monopole moment) such that the total convergence is the same \citep{Liesenborgs12}. The lens equation is unaltered outside of the circular region in question. Image positions, magnifications, and relative time delays are all recovered exactly. This transformation can be applied multiple times to different regions to drastically change the shape of the mass distribution with no affect on any lensing observable (example in Figure \ref{fig:monoexample}). Since this degeneracy does not directly affect time delays, it is more or less omitted from the discussion of $H_0$. 

Degeneracies need not reproduce the image positions exactly to have an effect on time delays. Observables can be reproduced well enough to be consistent with observations but not perfectly. \citet{Read07} showed that a general first-order perturbation to a power-law lens potential can produce zeroth-order changes in the time delay. It seems plausible that an imperfect monopole transformation could produce lenses which approximately reproduce image positions, but significantly alter time delays. Since the introduction of one or more monopoles to a lens can alter the shape of the density contours, model assumptions on the shape of the mass distribution may be tied to the monopole degeneracy in a similar way that model assumptions about slope are tied to the MSD (or SPT). Since no galaxy will perfectly match a given model, lensing degeneracies work behind the curtain of the fitting process to find a close degenerate solution within the assumptions of the model. Similar to how \citet{Gomer19} explored the effects of the MSD in the fitting process with slope as the focus, the effects of other degeneracies must be explored with the shape as the focus \citep{Saha06}, serving as partial motivation for this study.

\subsection{Ellipse+shear assumption}
Nearly all parametric lens models assume that the mass distribution is elliptical and has some external shear, which serves as a stand-in for external influences and higher order effects. For example, H0LiCOW uses a Bayesian inference between multiple models, but the primary model is their Singular Power Law Elliptical Mass Distribution model with external shear. While assumptions regarding the radial profile of lens mass distributions have been somewhat explored in the literature \citep{Enzi19}, this ubiquitous assumption about the azimuthal shape of the mass distribution has been less well-explored.

\citet{WW12} studied the azimuthal image positions of quad lenses, and found that for single-component elliptical mass distributions, the relative polar angles of the images lie on a well-defined surface, called the Fundamental Surface of Quads (FSQ). Because quads from all simple elliptical lenses, regardless of the degree of ellipticity or density profile shape, lie on the FSQ, it provides a model-independent benchmark for analyzing galaxy properties. Deviations from the FSQ signal departures from simple elliptical mass distributions. The observed galaxy-scale quad population has significant deviation from this surface, confirming that real lenses are not simply elliptical mass distributions. Even more interestingly, \citet{WW15} then showed that the addition of external shear was insufficient to bridge the gap, and that while many individual quads can be described as ellipse+shear, the population of quads cannot be reproduced-- it must come, at least in part, from lenses with more complicated mass distributions. \citet{Gomer18} expanded on this analysis, finding that the observed population cannot be accounted for by including $\Lambda$CDM substructure, even if the mass of each clump is increased by a factor of 10.

Images happen to lie at similar radii to the transition region from an inner component being baryon-dominated to an outer component being dominated by dark matter. It is quite likely that this transition region produces asymmetries to the lens shape.  \citet{Gomer18} went on to explore the effect of certain types of macro-structure within otherwise elliptical lenses. Two component profiles were constructed with a variety of perturbations added to the shape. The authors found that the only way to reproduce the observed azimuthal structure was through two-component profiles with a combination of Fourier components and offset centers, with a magnification bias present. This conveys that at least some fraction of mass distributions must be considerably more complicated than ellipse+shear models (even two-component models where the components are aligned) can describe. Because these mass models match the observed population in this respect, they are used as the starting point for the models in this work.

This is not to say that the ellipse+shear assumption has never been tested within the time-delay cosmography framework. H0LiCOW uses a two-component model as one of several models within a Bayesian framework, and this model has been used to introduce some level of asymmetry. However, most two-component models still assume that the centers of the two components coincide such that the overall shape is still elliptical. The baryon component may itself be composed of multiple components, eg. two Chameleon profiles. Some two-component models, such as the model of HE 0435-1223 by \citet{HC4}, also align centroids, but allow the position angles and ellipticities of the components to vary, providing a possible overall shape which is not elliptical. Again allowing the position angles (but not centroids) of the two components to vary, \citet{TDCOSMO1} found that a two-component model was able provide a better fit than a power law. However, the FSQ comparison of \cite{Gomer18} required a combination of perturbations to an elliptical shape beyond a difference in position angles, so even misaligned position angles such as in these models cannot provide sufficient asymmetry to match the observed quad population. While the main goal of H0LiCOW et al. is to apply models to fit specific lenses, one goal of this paper is to make sure that the lens models being prescribed are a good match to the general population of lenses.

Based on modeling of SLACS lenses, the two components of halos are not necessarily aligned, and instead can have different axis ratios or position angles. Offset centers of order 0.1 arcsec are not uncommon (see Figure 5 of \citet{Shajib19}). When a composite model is used in the H0LiCOW analysis of WFI2033-4723, the model center is offset from the light by $\sim 200$ pc \citep{HC12}. DM halos are not necessarily any more spherical than the light (see Table 1 of \citet{Shajib20b}). Since these measurements come from parametric fittings which assume all components are elliptical, they may not fully capture the inherent asymmetries of these systems. This paper seeks to explore a range of structure beyond a simple elliptical halo. 

Efforts to compare external shear with actual lens environments have found that in many cases the shear does not match what one would expect from the environment in either direction or magnitude \citep{Wong11}, implying that shear may not be a physical quantity as is typically assumed, but more of a first order fitting parameter which compensates for simplifying assumptions. \citet{Biggs04} used high resolution VLBA imaging to study a radio jet where three knots in the jet were multiply imaged. While a single knot could be fit with a Singular Isothermal Ellipsoid (SIE) + external shear model, it was not possible to fit all three images with SIE+shear. They were only able to fit the images by modeling the mass as a sum of Fourier components \citep{Evans03} which resulted in rather extreme ``wavy'' features. High resolution constraints using radio sources offer a testing ground for lens models, currently being explored by the strong lensing at high angular resolution program (SHARP). \citet{Spingola18} present their findings regarding the first target of the program, MG J0751+2716, wherein both a single-lens SIE+shear model and one accounting for the nearby galaxies drastically failed to reproduce image positions relative to the tight constraints of the VLBI observations. 

Coming from another angle, \citet{Night19} analyzed three SLACS lenses using PyAutoLens, a fully automated software which fits light and mass distributions simultaneously and determines the lens model complexity through Bayesian model comparison \citep{Night18}. The resulting models require two mass components which are offset both in terms of position angle and centroid position, effectively introducing a lopsidedness in the shape of the mass distributions, echoing the findings of \citet{Gomer18}.  Meanwhile, \citet{Williams20} also required a model with lopsided mass contours resulting from two offset mass components to fit the lensed supernova, iPTF16geu. \citet{Wagner19} developed a method to analyze lens systems in a model-independent way by comparing the observable properties of individual images locally to one another rather than globally to a particular model. The distinct separation of the locally-constrained regions near the images from the regions with no images where model assumptions are the only constraint allows a way to determine the effects of different model assumptions. The process makes it clear that only part of the information comes from the observational data alone-- a large part comes from the modeling assumptions. Comparing this method with others for the B0128+437 system, \citet{Wagner20} found that the lens could not be adequately fit as an ellipse+shear and that the implicit assumptions inherent to parametric modeling introduced incorrect local constraints which could not reproduce the millisecond image structure when applied globally.

The general trend seems to be that modifications to the ellipse+shear model are increasingly necessary as astrometry and modeling techniques improve. While the ellipse+shear model has been incredibly useful, it seems that it ignores (or possibly covers up) complexities in the mass distributions which are only now being revealed. 

Since the ellipse+shear assumption is not always representative of the true mass distribution, an analysis must be done to ascertain whether or not the assumption itself can introduce scatter or bias in the recovery of $H_0$, similar to how the assumptions on the radial profile shape can introduce biases through the MSD. The goal of this paper is to begin that discussion.

To explore the possible effects of this assumption, we will be producing quads from mock lenses which are not necessarily ellipse+shear, then fit the images as if they were real quads, with no knowledge of the true mass distribution, assuming ellipse+shear. Since there are many ways to construct a mass distribution and many ways to model lenses, the space of this problem has many facets to it. The context of this paper is to begin the exploration, but cannot comprehensively investigate all parts of it. Eventually, the effects of  spatially resolved kinematics, line of sight structure, and finite source size should all be included (as in H0LiCOW), but at present these are beyond the scope of this exercise. For now, we will simply try to match the image positions and time delays of these synthetic quads and determine the extent to which recovery of $H_0$ is affected.

\section{Template lens}\label{sec:template}

Before we create lenses with deviations from their simple elliptical shape, we need a control lens against which to compare our results. With the focus of this work being the effect of the ellipse+shear model, rather than the radial profile, we will use the same radial profile for all tests. A thorough exploration whether or not the ellipse+shear effect has any dependence on the radial profile (such as the presence of a core, or different DM concentration) is a task for future work. Our template mock lens will be a purely elliptical lens, such that the ellipse+shear model will be an accurate representation. As such, any bias or spread in the recovery of $h$, if present at all, would not be due to the ellipse+shear simplifying assumption. Later tests will use this lens as a template and add perturbations to the elliptical shape. 

This lens is constructed as a two-component potential with a steep power law component ($\alpha=0.4$, i.e. density slope $=-1.6$) representing baryonic matter and an NFW component representing dark matter ($r_{s}=10$ kpc). In addition to being physically motivated, the use of two components will eventually make for an elegant way to introduce perturbations from the elliptical shape by slightly changing one component relative to the other (Section \ref{sec:perturbations}). Since \citet{Gomer19} focused on the effects of radial profile shape while keeping the lens as ellipse+shear, this paper focuses on altering the azimuthal structure while keeping the radial profile intact. The lens we will use as a template is the same as the ``Model D'' lens from \citet{Gomer19}, who made four such model profiles from this formula. Of the four, this model is considered the most representative of real halos, as it has a slope near the image radius which is slightly steeper than isothermal \citep{slacskin3}.  The halo has a viral mass of $1.7\times10^{12}M_{\odot}$ and an Einstein radius of 5.5 kpc (0.82 arcsec), with dark matter becoming dominant at 2.0 kpc (0.30 arcsec), with a lens redshift of 0.6 and a source redshift of 3.0. Other physical attributes are available in Table 1 of \citet{Gomer19}. 

Lenses are created with $h_{input}$ = 0.7. Since we are interested in the bias of $h$, when recovered value of $h$ are quoted, they will be relative to 1.0, which corresponds to the correct recovery of $h$ = $h_{input}$.

\subsection{Fitting Procedure}
Each of our tests in this work will produce a set of lenses. One quad is generated for each lens by placing the source randomly within the caustic. Each of these quads is then fit as a power law using \texttt{lensmodel} \citep{Keeton01}. Detailed in \citet{Gomer19}, the procedure searches over 7 parameters: normalization, ellipticity, ellipse PA, shear, shear angle, core softening radius, and slope, minimizing $\chi^2$ and returning the best fit model and corresponding value of $h$.
% {\color{cyan} (with initializations drawn from the ranges listed in brackets): normalization, ellipticity $(1-q_{potential}) $[0,0.4] , ellipse position angle [$-90^{\circ}$,$90^{\circ}$], shear [0,0.4], shear angle [$-90^{\circ}$,$90^{\circ}$], core softening radius [0,0], and slope [-1.0,-1.35], minimizing $\chi^2$ and returning the best fit model and corresponding value of $h$. 
The search works through several steps. The first step holds all the parameters fixed except for mass normalization and position angles, and grid searches over both the ellipticity PA and shear angle. The second step frees ellipticity (defined as 1-$q_{potential}$) and shear, and grid searches over the interval [0,0.4]. The next step frees all the parameters and optimizes. The last step repeats the whole loop for a new initial value of slope. The core radius is initialized at zero, but can take on nonzero values in the second-to-last step, when all parameters are free to vary. 
By default, \texttt{lensmodel} uses a Gaussian prior for $h$, for which we set the sigma to be large ($10^6$) so as to effectively be a flat prior, as in the example in the \texttt{lensmodel} manual. Robustness of this fitting procedure is demonstrated in Appendix A of \citet{Gomer19}.

We fit the lens images in two different ways. In one fitting,  we fit the images as point sources with observational uncertainties of 0.003 arcseconds in spatial resolution and 0.1 days in time delays. \footnote{The astrometric errors are comparable with modern radio observations. The errors in time delay are likely more optimistic than present observations for quasar sources and are meant to be more forward-looking. (although they may be reachable for supernova sources, \citet{Wojtak19})} This fitting is the same as the fitting process in \citet{Gomer19}.  We hereafter refer to this fitting as the "point-source fitting".

In addition, we perform a second fitting in which the source includes an array of 9 point sources, meant to serve as an approximation for an extended source. These sub-sources are arranged in a cross configuration with two sources extending in each direction away from the center. Sub-sources are placed  at a distance of 0.05 and 0.1 arcseconds (0.1''= 0.67 kpc in the lens plane and 0.77 kpc in the source plane) from the center. The central point source has the same 0.003 arcsecond astrometric uncertainty, while the 8 sub-sources each have a positional uncertainty of 0.03 arcseconds (similar to the size of an HST pixel, for comparison a H0LiCOW image has a resolution of 0.05 arcseconds, \citet{HC4}). Only the central point source has a measured time delay, although we have relaxed the uncertainty on the time delay to 1 day. The logic behind this choice is to make the second fitting conform as closely to presently-observed lens systems as our framework allows, and 0.1 days is perhaps too optimistic. In addition, we are also curious to compare the effect of extended source information with that of enhanced time delay information- both constraints should help with degeneracies, but they may or may not eliminate similar sets of degenerate models. We hereafter refer to this fitting as the "extended-source fitting". Results for both fittings will be shared and compared.

\subsection{1ell test}
With the profile shape parameters set, mock quads can be created and fit. We create a population of 500 lenses from this profile shape by introducing ellipticity to the mass distribution. Axis ratios for the potential are uniformly chosen within the range of 0.85 to 0.99, which corresponds to roughly 0.5 to 0.99 with respect to mass (the same axis ratio is chosen for both the baryon and DM components). Lenses from this test have no complications to their elliptical shape.

This control test, designated ``1ell'', is the same test as that designated ``Model D'' in \citet{Gomer19}. Because the lens truly is an elliptical lens, the ellipse+shear model used to fit it is accurate. The only discrepancy between the created lens systems and the model used to fit them is that the model is a power law whereas the lens is a composite profile, a discrepancy explored deeply in \citet{Gomer19}. 

The fitting results are listed in Tables \ref{table:pointsource} and \ref{table:extsource}. The images are fit well, with $98\%$ ($99\%$) of quads having $\chi^2/dof<1$ for the point-source (extended-source) fittings.  Omitting the bad fits, the recovered ellipticity values correlate spectacularly with the input values ($R=1.00$ for the point-source fitting and $R=0.99$ for the extended-source fitting) and the recovered shear values are nearly zero (median = 0.0021 for the point-source fitting and 0.0053 for the extended-source fitting), indicating the fitting process has accurately matched the true mass distribution shape quite well. With well-recovered ellipticities and shears, the fitting procedures have not been strongly influenced by the ellipticity-shear degeneracy.

\begin{figure*}
 \centering
   \includegraphics[width=\linewidth]{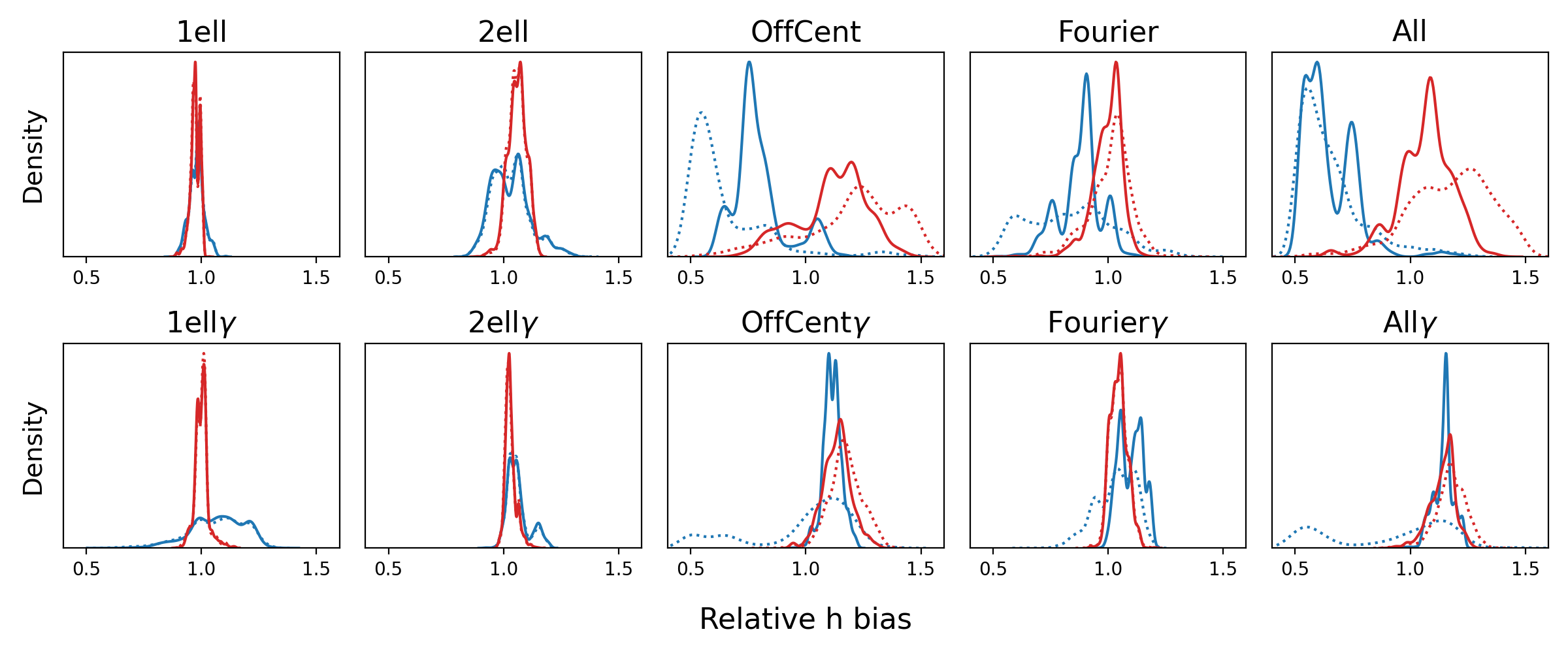}
   \caption{The posterior distributions of $h$ for each test for the point-source fitting (blue distribution) and the extended source fitting (red distribution). When quads with $\chi^2/dof>1$ are omitted, the solid curves are recovered, while the whole sample with no selection corresponds to the dotted curves. The distribution is estimated by bootstrapping the set of the combined MLE determination of $h$.}
\label{fig:halltests}
\end{figure*}

The recovered distribution of $h$ is depicted in the upper left panel of Figure \ref{fig:halltests}. Each lens recovers a single best-fit value of $h$, but it is more useful to consider the value one would get by combining the fits together, as is done in studies of real systems such as H0LiCOW. To represent this value, a Maximum Likelihood Estimation (MLE) is performed by taking the distribution of $\chi^2/dof$  with respect to $h$ near the best-fit value and calculating a likelihood as a function of $h$. The other parameters are marginalized over for this calculation. Likelihoods are calculated for each lens. The lenses with $\chi^2/dof<1$ best fits have their likelihoods combined together. To accurately determine scatter, the distribution is bootstrapped and shown in Figure \ref{fig:halltests}. The mean and standard deviation of the $h$ distribution is listed in Table \ref{table:pointsource}.

For ``1ell'', the recovered value of $h$ relative to 1.0 is $0.98\pm0.03$ for the point-source fitting ($0.98\pm0.02$ for the extended-source fitting). The scatter of the distribution is such that it is consistent with being unbiased. This distribution of $h$ will be the standard against which later tests will be compared. If changes to the elliptical shape cause the recovered value of $h$ to drastically change from this value, we can conclude that the ellipse+shear assumption has a biasing effect on $h$.

\section{Addition of perturbations}\label{sec:perturbations}
At the heart of this exercise is the question of why the ellipse+shear model may not be sufficient. Perhaps the most plausible physical motivation for perturbations to the elliptical shape of halos comes from noting that the image radius ($R_E=5.5$ kpc in the circular case) is not too far-removed from the transition radius where dark matter begins to dominate the mass ($R_{trans}= 2.0$ kpc again in the circular case for the lens template). If the baryon component and the dark matter component have even slightly different shapes, the mass distribution in this transition region will not have purely elliptical contours. Slight adjustment of the alignment of the two components offers a natural way to make the lens slightly non-elliptical, an effect which \citet{Shajib19} measure in some real lenses (see Fig. 5 therein).

\citet{Gomer18} discovered that only these types of perturbations to the elliptical shape were capable of reproducing the angular distribution of quad images. As such, we seek to reproduce the types of perturbed shapes which \citet{Gomer18} found to be necessary. Several types of alterations to the elliptical shape were used to introduce slight asymmetries. These include having different ellipticities for the two mass components, misalignment of the position angles and offset centers for the two components, and the addition of $a_4$ and $a_6$ Fourier components \citep{Bender87}. \citet{Gomer18} found a combination of these perturbations to be necessary to explain the population, so it will be necessary to examine them individually and collectively in a series of tests, which are detailed below.

Unlike ``1ell'', all further tests  have position angles for the baryons and dark matter components which are offset by 15 degrees.
While \citet{Gomer18} allowed this angle offset to vary between 0 and 45 degrees, we restrict it to a nonzero value in the interest of controlling variables. Whereas ``1ell'' kept the ellipticity the same for both components, observed lenses appear to commonly have different axis ratios for the two components \citep{Shajib19}, and so we have set both components to have separately drawn axis ratios which range (in potential) from 0.85 to 0.99.

The following list details additional complications specific to each test. Values are chosen to match the shape perturbations discussed in \citet{Gomer18}:
\begin{itemize}
 \item 2ell- No further complications are present aside from the misaligned axes and separate ellipticities of the two components. 
 \item OffCent- The centers of the mass distributions are offset by up to 1 kpc in a random direction to introduce lopsidedness. Because the offset coordinate is distributed uniformly by radius, the offsets are more centrally concentrated than a uniform distribution within the area. The \texttt{lensmodel} fit fixes the center of the mass distributon to the center of the baryon distribution. The offset center can be thought of as introducing a mass dipole moment around the center of light of the lens.
 \item Fourier- Centers are coincident, but Fourier components are added, with $a_4$ in the range of [-0.005,0.005] and $a_6$ in the range of [-0.001, 0.001] with respect to potential\footnote{
        The listed Fourier values are the coefficients in front of the cosine term. In the notation of \citet{Xu15}, which explicitly includes a normalization for the multipole moment, these values correspond to $a_4 \in [-0.075,0.075]$ and $a_6 \in [-0.035,0.035]$  }.   
The resulting mass distributions visually match the same range as \citet{Gomer18}, but the values are different since previously components were added with respect to mass.
 \item All- The centers of the mass distributions are offset by up to 1 kpc as in ``OffCent''. Fourier components are added, with the same range of values as ``Fourier''.
\end{itemize}

\begin{figure}
 \centering
   \includegraphics[width=\linewidth]{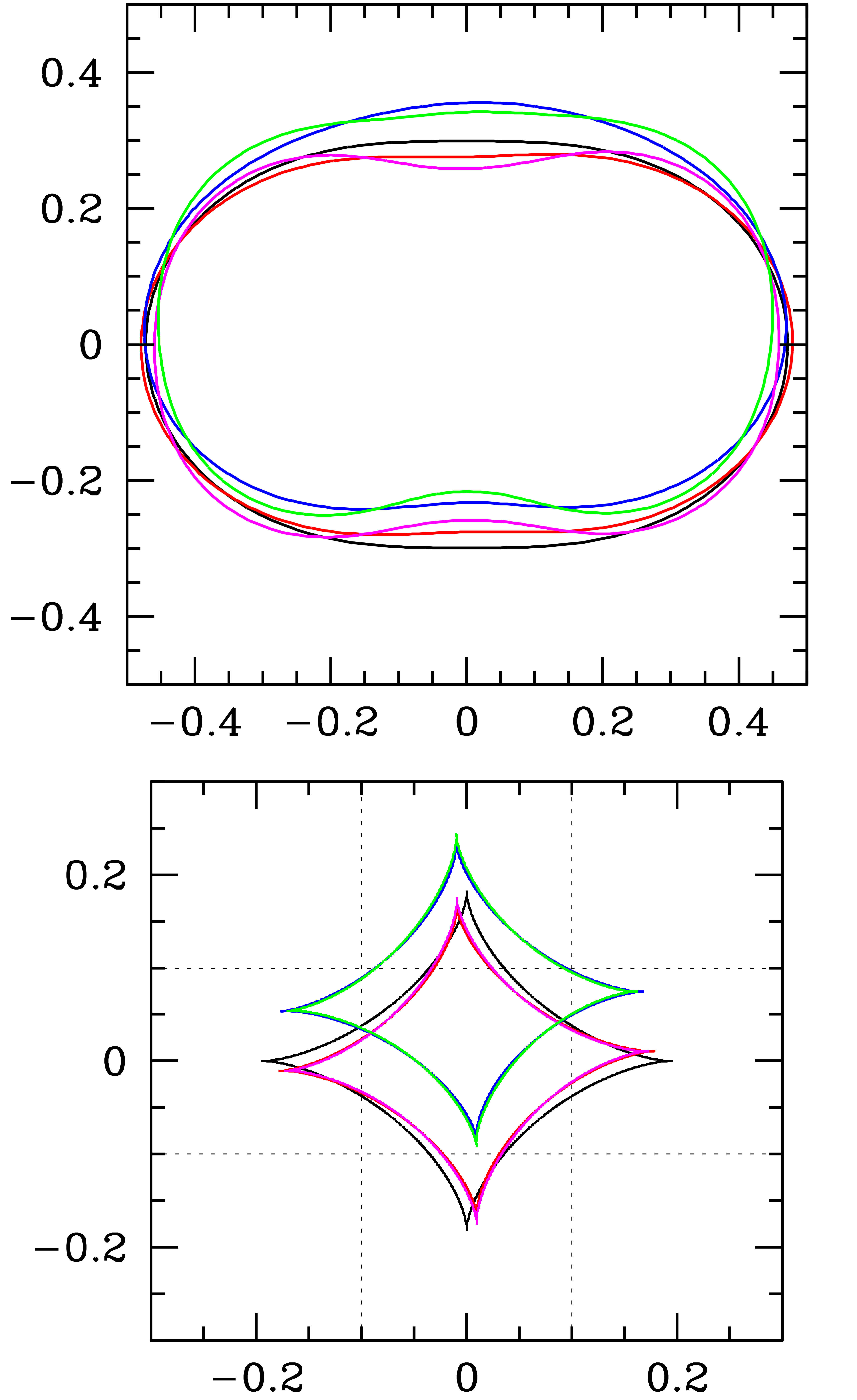}
   \caption{The $\kappa=1$ mass density contour and caustic for each of the 5 types of tests, shown to illustrate the nature of each type of perturbation to the elliptical shape.  The degree of extremity for all these perturbations is set to its highest level in this example-- all created lenses for the tests in this paper have perturbations of equal or lesser magnitude to this example. Black corresponds to ``1ell'', where the lens is a single ellipse. The other four tests use different axis ratios for the two components with one major axis tilted by 15 degrees relative to the other. Red corresponds to ``2ell''. Blue corresponds to ``OffCent'', where one center is offset, in this case 1 kpc upward. Magenta corresponds to ``Fourier'' where both components have Fourier perturbations added. Finally, the green contour and caustic correspond to the ``All'' test, with all of the above perturbations included simultaneously.}
\label{fig:testshape}
\end{figure}

As an illustration of each type of perturbation, Figure \ref{fig:testshape} shows a single $\kappa=1$ mass density contour for an extreme example lens from each test. Caustics are also shown. Generally, the misalignment of the position angles of the two components slightly tilts the caustics of the other tests compared to ``1ell'', with only subtle changes to the mass contour at $\kappa=1$. The offset centers of ``OffCent'' and ``All'' cause
the mass contour to be lopsided, while also displacing and slightly deforming the caustics. Meanwhile, the Fourier perturbations of ``Fourier'' and ``All'' add wavy features to the shape of the mass contours but do not significantly alter the caustics.

In addition to these five tests, we also wish to include tests with external shear. Significant external shear is necessary to reproduce the ratio of radial image positions in observed quad systems, which will be discussed in detail in Section \ref{ssec:obscompare}. In addition to the above tests, we also run five tests with external shear included in lens construction.

\begin{itemize}
 \item 1ell$\gamma$- Same as ``1ell'' in that position angles are aligned and axis ratios identical for both components. A randomly oriented shear is introduced with $\gamma$ between 0 and 0.4.
 \item 2ell$\gamma$- Same as ``2ell'' except that a randomly oriented shear is introduced with $\gamma$ between 0 and 0.4.
 \item OffCent$\gamma$- Same as ``OffCent'' except that a randomly oriented shear is introduced with $\gamma$ between 0 and 0.4.
 \item Fourier$\gamma$- Same as ``Fourier'' except that a randomly oriented shear is introduced with $\gamma$ between 0 and 0.4.
 \item All$\gamma$- Same as ``All'' except that a randomly oriented shear is introduced with $\gamma$ between 0 and 0.4.
\end{itemize}

For each test, 500 lenses are created, each producing a single quad, which is fit with the same \texttt{lensmodel} routine as before. The resulting recovery of $h$ for each test is depicted in Figure \ref{fig:halltests}, with the results in Table \ref{table:pointsource}.

\subsection{Limitations}\label{ssec:limitations}
The main advantage of this study is that the deviations from a simple elliptical shape are known beforehand and controlled to each different test. However, there are some disadvantages that come from this type of study as well. For example, stellar kinematics are commonly used to break degeneracies through the spherical Jeans approximation \citep{HC4}. In our case, it would be difficult to know the effect of our complications on the velocity dispersion. Because we lack adequate mock kinematics, the stellar kinematic information which is used to break the MSD has not been included in this fitting process. The role of stellar kinematic constraints was discussed by \citet{Gomer19}, who found that models for kinematics which do not match the lens exactly can cause an incorrect breaking of the MSD, leading to bias. The decision to omit stellar kinematics from this paper is an attempt to control different forms of bias, although in future work these simplifying assumptions will need to be considered in aggregate.

In real systems, there is information to be gained from the ring resulting from extended sources, which is used to help control degeneracies. The extent to which this can help is debated \citep{Saha01,Walls18,HC1}, and may be subject for further exploration. We have attempted to address this by supplementing our point-source fitting with the second fitting which uses an array of sources. This still may not fully capture the information from an extended source, but it allows us to analyze more quad systems since the lenses are simpler to synthesize and fit.

% Meanwhile, especially for the point-source fittings, we have set optimistic error bars on these values, emulating very good astrometry and time delay recovery. The quality of these mock observations may help offset some of the other limitations above.

\subsection{Comparison with observed quads}\label{ssec:obscompare}
Before we draw any conclusions about the recovery of $h$ for these quads, we must first confirm that the population of mock quads is representative of the observed population. Only then can we be confident that our results will be generally applicable to real lens systems. 

Rather than comparing properties recovered from modeling, such as ellipticity or shear, we would like to compare quad populations independent of the modeling process. As such, we look at statistical distributions of image properties, namely the distribution of relative image angles relative to the FSQ \citep{WW15} and the radial distance ratios. Any set of mock quads which seeks to represent a real population should at the minimum match the statistical properties of the observed population of quads.

The first statistical comparison to make is to compare the distributions of the angular positions of the images. Four images are uniquely defined by three relative angles, which can be plotted in 3D space. When plotted in this space, quads which come from elliptical mass models will lie on the FSQ \citep{WW12}. Meanwhile, the distribution of observed quads has significant spread from the FSQ. External shear causes the distribution to split above and below the FSQ, but is insufficient to account for the observed distribution \citep{WW15}. However, the perturbations to the potential we use in our various tests can reproduce the observed distribution \citep{Gomer18}. 

These perturbations were sufficient to recreate the observed distribution when used in conjunction with a magnification bias, which selectively removes systems with lower magnifications from the comparison set with a probability of being kept being proportional to the summed magnifications. The bias is meant to emulate the fact that brighter systems are more likely to show in surveys, and therefore our data set, while dim systems are more likely to be overlooked. For real systems, the probability of being found is also related to the brightness of the quasar, but since this is independent of the lensing effect, the magnification of the images is the relevant parameter to describe this effect. For consistency, when we compare our mock quads with the observed population, we will also apply this magnification bias to our population of quads.

\begin{figure}
 %\centering
\includegraphics[width=0.97\linewidth]{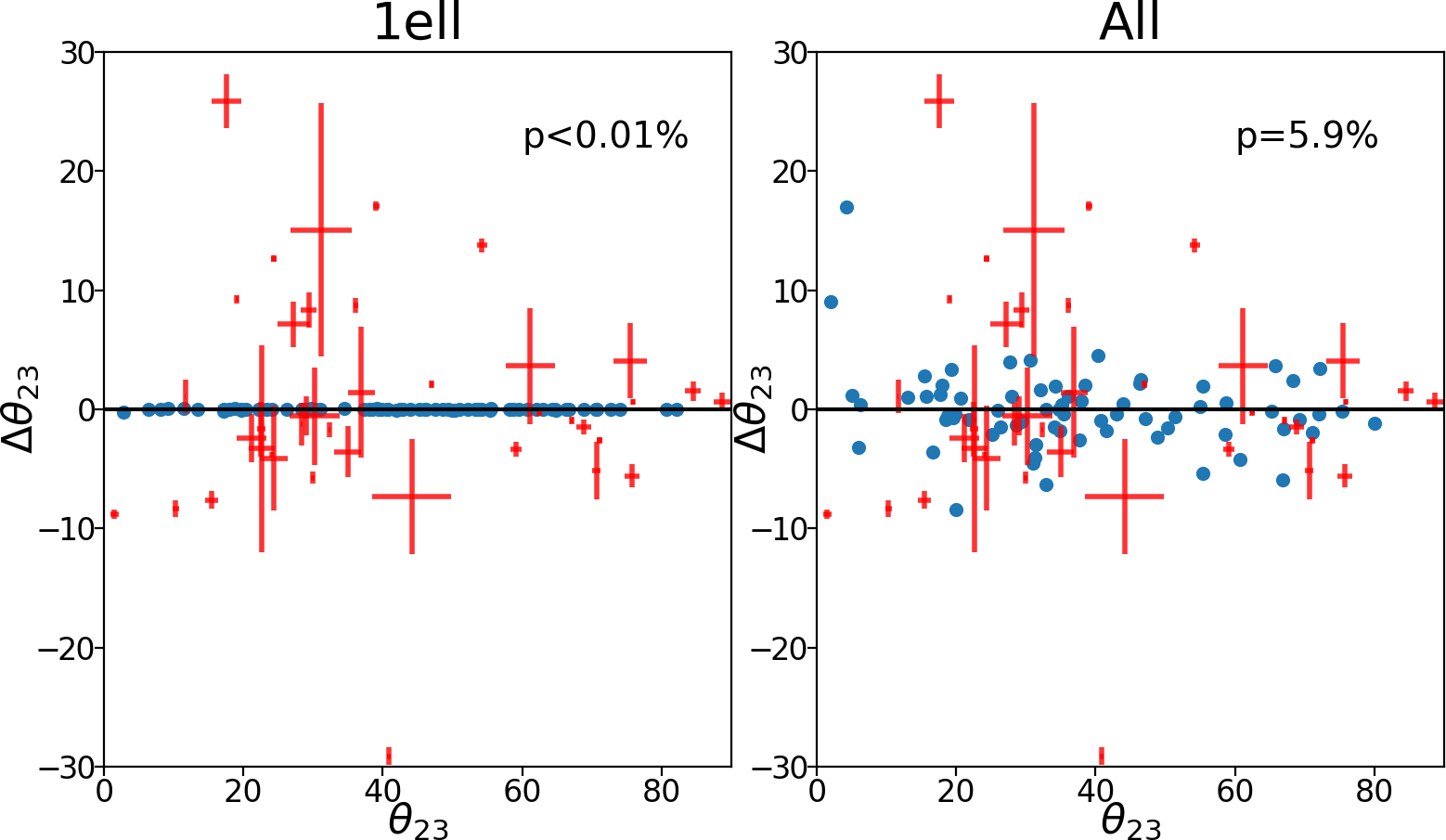}
\caption{ 
      The distribution of quads relative to the Fundamental Surface of Quads, projected such that deviation from the FSQ is the vertical deviation from zero. The observed quad population (red) has considerable deviation from the FSQ (and therefore an elliptical mass profile). The blue points represent synthetic quads from the ``1ell'' test (left) and the ``All'' test (right). The  ``1ell'' quads do not deviate from the FSQ and are not consistent with the observed population, while the ``All'' quads are consistent with the observed population in this respect.
           }
\label{fig:FSQcompare}
\end{figure}

Previous work comparing the observed population to the FSQ provides evidence for the presence of azimuthal substructure in galaxy lenses and was the motivation for this work. As such, we should confirm that the quads involved match the observed population in this context. Figure \ref{fig:FSQcompare} shows the distribution of quad image angles relative to the FSQ for ``1ell'' (pure ellipse) and ``All'' (all types of perturbations added). While ``1ell'' fails to reproduce the scatter relative to the FSQ, ``All`` is more consistent with the observed population in this respect, with a p-value of $\simeq5\%$, replicating the findings from \citet{Gomer18}. 

\begin{figure}
\centering
\includegraphics[width=0.95\linewidth]{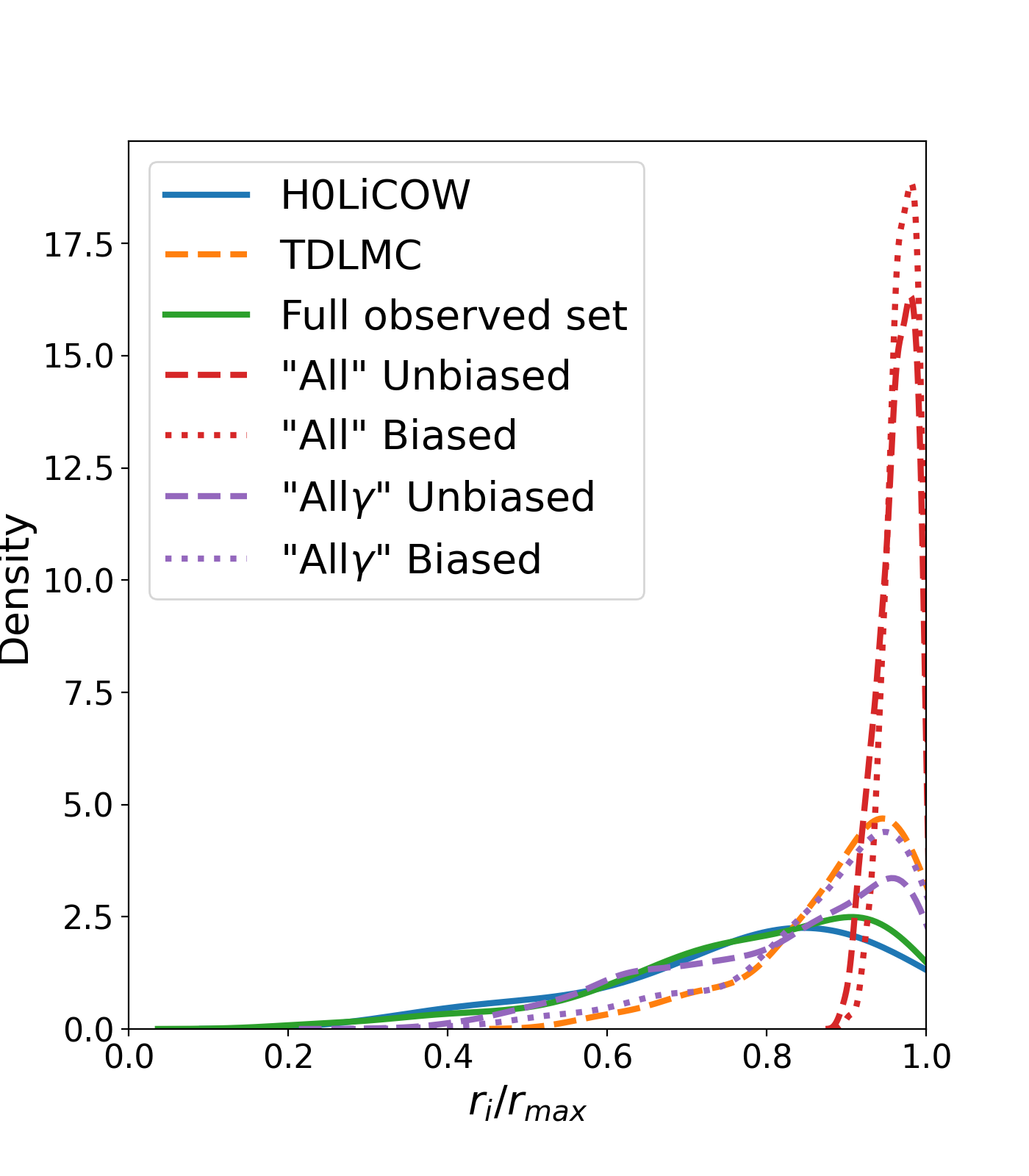}
\caption{ 
    The radial distribution of images for quads as a ratio to the farthest-out image. The H0LiCOW quads (solid blue) are a subset of the observed quads used in \ref{table:quadlist} (solid green). The ``All'' (red) and ``All$\gamma$'' (purple) tests from this work are shown, both as a full set (dashed) and with a selection bias applied based on the summed magnification (dotted). The synthetic Rung 3 TDLMC quads (yellow dashed line) are also plotted. The clear discrepancy between the ``All'' test and the observed quads illustrates the need for some additional perturbation to match the observed set. The large shear in ``All$\gamma$'' helps to better match the distribution. The biased ``All$\gamma$'' population returns a p-value of 3.1\% when compared to the H0LiCOW population, higher than the ``All'' ($<1\%$) or the TDLMC (2.1\%) populations) Note that our biased ``All$\gamma$'' set is very similar to the TDLMC numerically simulated quads. }
    
\label{fig:imageratios}
\end{figure}

In this work we seek to expand the model-free population comparison toolkit by adding a comparison with respect to image distance ratios relative to the lens center. Figure \ref{fig:imageratios} shows the observed distribution of radial image ratios as solid curves (the full set of quads used in this work listed in Table \ref{table:quadlist} in green and the H0LiCOW subset in blue). Relative to the farthest-out image, the observed population of quads consists of a large spread of image distances, ranging from quads with multiple images at approximately the same image radius ($r_i = r_{max}$), to quads with some images drastically closer to the center than the outermost image ($r_i = 0.3r_{max}$, for example). In the same figure, we plot two synthetic sets from this work: the ``All'' and ``All$\gamma$'' tests. Clearly the ``All'' quads (red distribution) are a poor representation of real systems in this respect, as the images lie at too similar of radii, even with the additional perturbations to the ellipse+shear model. It appears very difficult to get the image distance ratio distribution to be as broad as in observations. This is not unique to our quads, and in fact appears to be a problem with quads from numerical simulations, or at least those with a single lens plane and located in simple environments.

The Time Delay Lens Modeling Challenge (TDLMC, \citet{Ding18}) has created a population of synthetic quads to be fit and modeled, serving as a standard to compare the accuracy of time delay lens models. The synthetic quads come from 3 rungs, with the first 2 sets being created as elliptical power law lens models \citep{Ding21}. Rung 3 was constructed from numerical simulations, namely Illustris \citep{Vogelsberger14} and the zoom-in simulations of \citet{Frigo19}. The time delay results of Rung 3 are difficult to interpret due to numerical resolution effects, but since they are created from numerical simulations and are therefore more complicated than a simple ellipse+shear model, we also plot the Rung 3 lenses as the yellow dashed curve in Figure \ref{fig:imageratios}. We can see that the TDLMC Rung 3 quads also do not span the same range of radii as the observed population.

The second set of tests with shear (with the $\gamma$ suffix) was created to attempt to address this concern. The most straightforward way to create quads with smaller image ratios is to introduce significant amounts of shear. As such, these tests are the same as the first set but with shear introduced between 0 and 0.4, with the range chosen simply to more closely match the image ratios (purple distributions in \ref{fig:imageratios}). While these values of shear may seem extreme, it is important to restate that external shear likely represents more of a fitting parameter than a physical quantity \citep{Wong11}. The creation of these tests is not a claim that lens environments produce physical shears which are this large, but rather is just another perturbation added to the quad creation to attempt to recreate the statistical population of quad images. As is commonly done in lens modeling, shear serves as a first-order approximation for many types of perturbations. It is conceivable that the reason real lenses have such extreme image ratios is some unknown perturbation to their shape which we have not included, but shear serves to approximate. The important takeaway from this is that the tests with $\gamma$ generate quads which provide a closer match to the observed distance ratios than those without $\gamma$.

All told, some of the 10 tests are more capable of matching certain characteristics of the observed quad population than others. The ``All'' test best matches the angular distribution, while the tests with $\gamma$ attempt to match the observed radial distribution of images. However, once the same magnification bias as \cite{Gomer18} is applied, even the $\gamma$ tests with a shear ranging between 0 and 0.4 are still unable to match the observed distribution of distance ratios. The $p$ value for comparing the ``All$\gamma$'' case with the magnification bias to the for the H0liCOW set is still 3.1\% ($<1\%$ for the full observed set). The inability of this lens population to fully account for the observed radial spread of images reflects a deep mystery about the structure of lens galaxies, which will be discussed further in Section \ref{ssec:radratiocauses}. The distribution closely matches the TDLMC Rung 3 distribution in Figure \ref{fig:imageratios} (with a slightly better $p$ than the TDLMC quads), perhaps indicating that it is similarly comparable to the observed quads. We continue to use the $\gamma$ test populations as a comparison, noting that this inadequacy is no worse for our set than any synthetic population currently available.

\begin{figure}
 \centering
 \includegraphics[width=\linewidth]{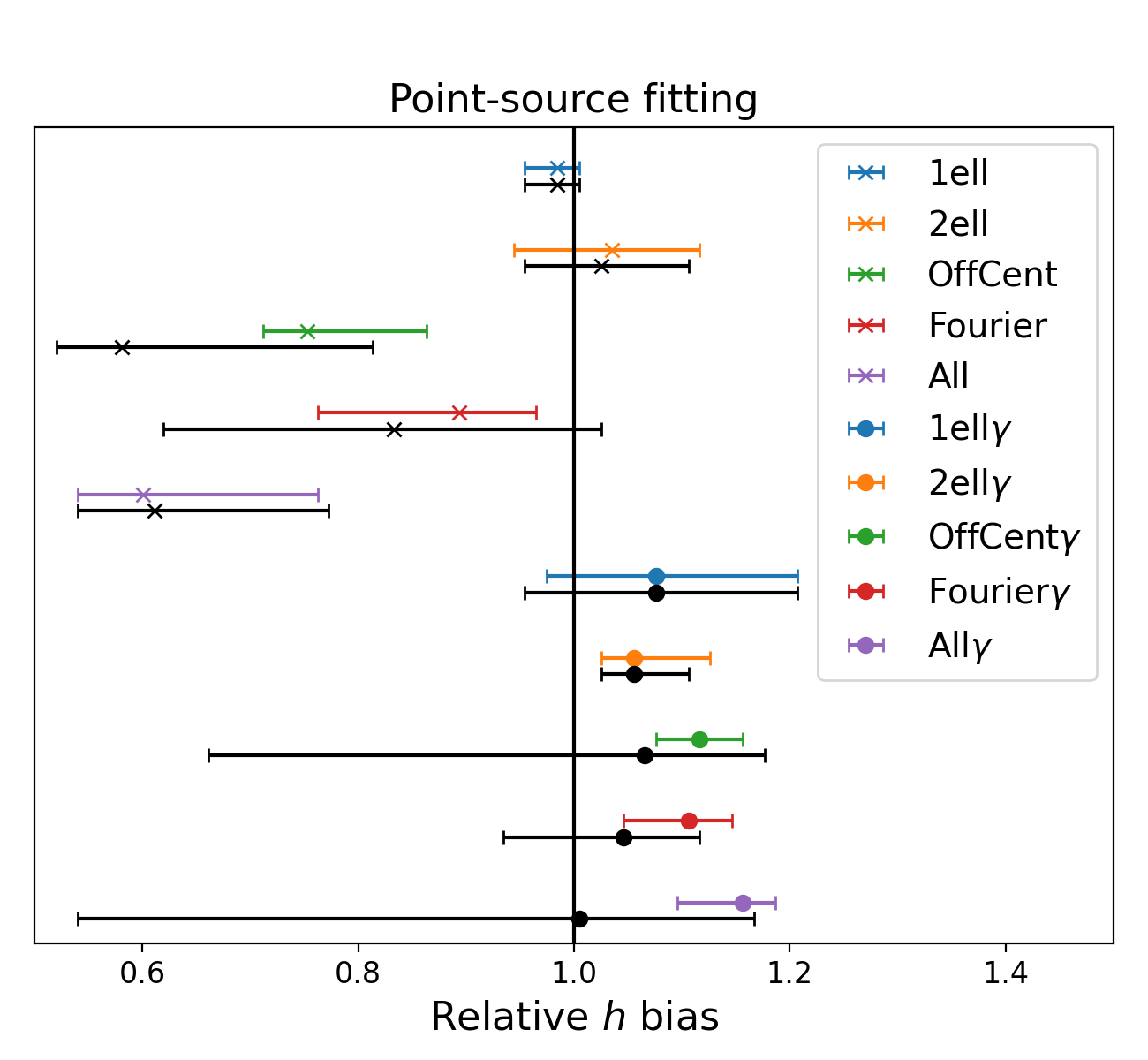}
 \includegraphics[width=\linewidth]{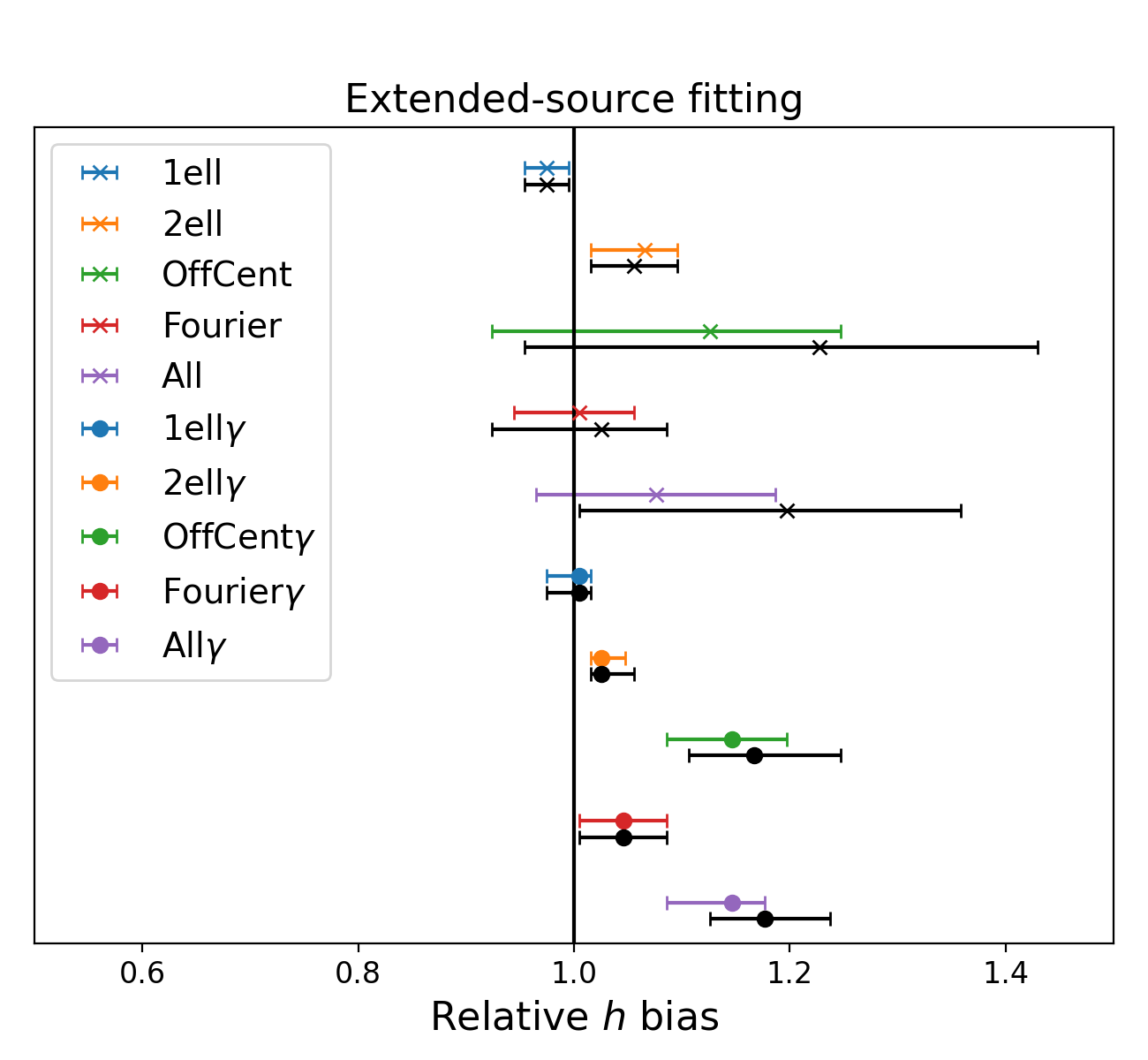}
    \caption{ A side-by-side comparison of the recovered distribution of $h$ for each of the tests in this paper (top: point-source fitting, bottom: extended-source fitting). For each test, the colored bar refers to the 68\% confidence interval of the good sample, omitting quads with $\chi^2/dof<1$, while the black bar below each colored bar depicts the result if the whole sample is used. 
           }
\label{fig:hwhisker}
\end{figure}

\section{Results}\label{sec:results}

The results from the 10 tests in this paper are described in Tables \ref{table:pointsource} and \ref{table:extsource}, with the recovered distributions of $h$ in Figures \ref{fig:halltests} and \ref{fig:hwhisker}.

We define the subset of systems with $\chi^2/dof<1$ as the good sample, as opposed to the whole sample of all 500 quads for each test. We list the fraction of systems which are fit with $\chi^2/dof<1$ for each test in each table. Unless otherwise stated, we will refer to the result of the good sample, although the whole sample will be discussed in Section \ref{ssec:hramifications}.

\subsection{Fitting success}
Before discussing $h$, we wish to draw attention to our measures of how successful we are in reproducing the lens mass distribution features. Working with only the good sample, we consider two measures to check if the values of ellipticity and shear are correctly recovered: the correlations between the values used in lens construction to the recovered values. 

Perhaps unsurprisingly, the ``1ell'' test has a perfect correlation between the input ellipticity and the fit value, as well as a high fraction of quads which are successfully fit with good $\chi^2/dof$. This is expected because for this test the ellipse+shear model accurately describes the lens. Similarly, the ``1ell$\gamma$'' test successfully fits nearly all quads and recovers the parameters accurately because the model matches the lens.

The other tests have varying degrees of success. The ``2ell'' and ``2ell$\gamma$'' tests successfully recover the ellipticity and shear of their lenses, despite the dark matter axis being tilted with respect to the baryon distribution. Fourier components seem to only marginally affect this correlation success, but the tests with Fourier components fail to fit the majority of their quads with $\chi^2/dof<1$. Tests with offset centers fail on both accounts-- most fits have $\chi^2/dof>1$ and the recovered ellipticity is poorly correlated with the actual value. The ``All'' and ``All$\gamma$'' tests fall into the same category. For all tests with external shear, the recovered shear correlates strongly with the input value, likely because it is the same type of perturbation as anticipated in the model.

For the extended source fitting (Table \ref{table:extsource}), the measures of success are similar, although in general the fits across all tests have better $\chi^2/dof$. More systems are successfully fit with $\chi^2/dof>1$, a majority in each test but the "All" case. This better $\chi^2/dof$ comes primarily from a larger number of degrees of freedom, with $\chi^2$ alone being similar in both fittings. Despite more good fits, the correlations with respect to ellipticity and shear are mostly unchanged.

Two of these measures of success are observable for a population of real systems, namely the correlation between baryon and mass model ellipticity as well as $\chi^2/dof$. For example, it is interesting that the ``2ell'' tests retain a good ellipticity correlation, because this implies that real dark matter halos could have misaligned axis ratios without providing an observable effect on the ellipticity correlation. That is, just because the baryon elliptical axis is strongly correlated with the mass model, it does not necessarily mean that the two components are aligned or have the same axis ratios. Offset centers of the two distributions affect these quantities most drastically, so if real lenses have poor fits or ellipticity correlations, then offset centers could be an explanation. For comparison, results from the EAGLE simulation in \citet{Tagore18} have only about $48\%$ of quads fit with $\chi^2/dof<1$, perhaps indicating complications to the ellipse+shear model. This fraction for our tests is listed in Tables \ref{table:pointsource} and \ref{table:extsource}.

% \begin{landscape}
{\renewcommand{\arraystretch}{1.5}
\begin{table*}

\centering
\setlength\tabcolsep{3pt}
\begin{tabular}{c c c c c c c c c c c}
\multicolumn{11}{c}{Point-source fitting results} \\
\hline
Test & Shear? & Different q? & Tilted PA? & Fourier? & Offset centers? & Good sample $h$ & Whole sample $h$ & Ell $R_{corr}$ & $\gamma$ $R_{corr}$ & $f_{\chi^2/dof<1}$ \\
1ell & -- & -- & -- & -- & -- & $0.98^{+0.02}_{-0.03}$ & $0.98^{+0.02}_{-0.03}$ & 1.00 & -- & 0.98 \\ %n=50
2ell & -- & X & X & -- & -- & $1.04^{+0.08}_{-0.09}$ & $1.03^{+0.08}_{-0.07}$ & 0.93 & -- & 0.99  \\ %n=50
OffCent & -- & X & X & -- & X & $0.75^{+0.11}_{-0.04}$ & $0.58^{+0.23}_{-0.06}$ &  0.34 & -- & 0.21  \\ %n=50
Fourier & -- & X & X & X & -- & $0.89^{+0.07}_{-0.13}$ & $0.83^{+0.19}_{-0.21}$  &  0.83 & -- & 0.09  \\ %%less than 100 quads, only 22 used for bootstrap
All & -- & X & X & X & X & $0.60^{+0.16}_{-0.06}$  & $0.61^{+0.16}_{-0.07}$  &  0.35 & -- & 0.05  \\ %%less than 100 quads, only 11 used for bootstrap
1ell$\gamma$ & X & -- & -- & -- & -- & $1.08^{+0.13}_{-0.10}$  & $1.08^{+0.13}_{-0.12}$ &  1.00 & 1.00 & 1.00 \\ %n=50
2ell$\gamma$ & X & X & X & -- & -- & $1.06^{+0.07}_{-0.03}$  & $1.06^{+0.05}_{-0.03}$ & 0.91 & 0.99 & 0.98  \\ %n=50
OffCent$\gamma$ & X & X & X & -- & X & $1.12^{+0.04}_{-0.04}$ & $1.07^{+0.11}_{-0.40}$ & 0.59 & 0.92 & 0.31 \\ %n=50
Fourier$\gamma$ & X & X & X & X & -- & $1.11^{+0.04}_{-0.06}$ & $1.05^{+0.07}_{-0.11}$ &  0.83 & 0.95 & 0.35 \\ %n=50
All$\gamma$ & X & X & X & X & X & $1.16^{+0.03}_{-0.06}$ & $1.01^{+0.16}_{-0.46}$ & 0.47 & 0.86 & 0.17 \\ %%less than 100 quads, only 42 used for bootstrap

\end{tabular}
\caption{
    A summary table of the tests in this paper. For each test, the type(s) of perturbations to the ellipse+shear model are designated. For each test, the MLE determination of $h$ is shown, as well as the measures of fitting success: the Pearson correlation coefficients between input and recovered ellipticity and shear, and the fraction of systems which were successfully fit with $\chi^2/dof<1$. The good sample refers to only those with $\chi^2/dof<1$, while the whole sample refers to all quads. Correlations are calculated for the good sample. Recovered values of $h$ are relative to an unbiased value of 1. }
    % The last column is the combined p-value of the distributions of relative distance ratios and angular image coordinates of each test compared to the observed population.}
\label{table:pointsource}
\end{table*}}
% \end{landscape}

{\renewcommand{\arraystretch}{1.5}
\begin{table*}
\centering
\setlength\tabcolsep{3pt}
\begin{tabular}{c c c c c c c c cc c}
\multicolumn{11}{c}{Extended-source fitting results} \\
\hline
Test & Shear? & Different q? & Tilted PA? & Fourier? & Offset centers? & Good sample $h$ & Whole sample $h$ & Ell $R_{corr}$ & $\gamma$ $R_{corr}$ & $f_{\chi^2/dof<1}$\\
1ell & -- & -- & -- & -- & -- & $0.98^{+0.02}_{-0.02}$ & $0.97^{+0.02}_{-0.02}$ & 0.99 & -- & 0.99 \\ 
2ell & -- & X & X & -- & -- & $1.07^{+0.05}_{-0.03}$ & $1.06^{+0.04}_{-0.04}$ & 0.94 & -- & 1.00  \\ 
OffCent & -- & X & X & -- & X & $1.13^{+0.20}_{-0.12}$ & $1.23^{+0.27}_{-0.20}$ &  0.62 & -- & 0.49  \\ 
Fourier & -- & X & X & X & -- & $1.01^{+0.06}_{-0.05}$ & $1.03^{+0.10}_{-0.06}$  &  0.81 & -- & 0.71  \\ 
All & -- & X & X & X & X & $1.08^{+0.11}_{-0.11}$  & $1.20^{+0.19}_{-0.16}$  &  0.57 & -- & 0.38  \\ 
1ell$\gamma$ & X & -- & -- & -- & -- & $1.01^{+0.03}_{-0.01}$  & $1.01^{+0.03}_{-0.01}$ &  1.00 & 1.00 & 1.00 \\ 
2ell$\gamma$ & X & X & X & -- & -- & $1.03^{+0.01}_{-0.02}$  & $1.03^{+0.01}_{-0.03}$ & 0.93 & 0.99 & 1.00  \\ 
OffCent$\gamma$ & X & X & X & -- & X & $1.15^{+0.06}_{-0.05}$ & $1.17^{+0.06}_{-0.08}$ & 0.45 & 0.84 & 0.79 \\
Fourier$\gamma$ & X & X & X & X & -- & $1.05^{+0.04}_{-0.04}$ & $1.05^{+0.04}_{-0.04}$ &  0.77 & 0.95 & 0.94 \\ 
All$\gamma$ & X & X & X & X & X & $1.15^{+0.06}_{-0.03}$ & $1.18^{+0.05}_{-0.06}$ & 0.42 & 0.82 & 0.73 \\ 

\end{tabular}
\caption{Same as Table \ref{table:pointsource} but for the extended-source fittings. 
   }
\label{table:extsource}
\end{table*}}

\subsection{Recovery of h}
We now turn to the recovery of $h$ for the different tests. The bootstrapped MLE distributions are shown in Figure \ref{fig:halltests}, with the median and $1\sigma$ errors in Table \ref{table:pointsource}. The question of interest is to what extent the introduction of shape perturbations has changed the recovery of $h$. To this end, the $h$ distributions of the other tests should be compared to the ``1ell'' test, which has a simple elliptical lens. 

We stress that because only the fits with good $\chi^2/dof$ are included, the results are what one would get from fitting these systems in reality, ignorant of the complications to the shape of their true mass distributions. Even though these results all have $\chi^2/dof<1$, some systems recover very biased values of $h$. Since the fits are good, these cases would likely not raise any notice if they were real systems, which could introduce untrustworthy results. Perhaps the only measurable indication one would have is that only a fraction of systems within a large population are successfully fit with $\chi^2/dof<1$ when an automated fitting procedure is uniformly applied.

\subsubsection{Point source fitting}

The ``1ell'' test result is what one would hope from this kind of analysis. The MLE combination results in a constraint with $1.5\%$ bias downward and $3\%$ scatter. This is not precise enough for a $1\%$ determination, but it it is consistent with an unbiased recovery of $h$ and serves as a good point of comparison for the effects of perturbations to the elliptical shape. 

The resulting $h$ of all 10 tests can be compared with one another in Figure \ref{fig:hwhisker}. The ``2ell'' test also resulted in recovered values of $h$ consistent with the unbiased case, although the scatter has increased considerably ($\sim10\%$, with the median at 1.04). The ``Fourier'' test has similar scatter, although the result is biased at just over $1\sigma$, with the median at 0.89. The tests with offset centers, ``OffCent'' and ``All'', result in particularly bad recoveries of $h$, biased downward by 25 and 40 percent, respectively, with non-Gaussian scatter of order $10\%$. The ``All'' test was the most extreme of these five, but it was also the only one which matched the angular distribution of images. 

The $\gamma$ tests have large external shear in the lenses, which has changed the result considerably. All five of these tests return approximately the same median value of $h$, biased $\sim10\%$ above the true value. The scatter for all of these tests, except ``1ell$\gamma$'', has decreased compared to their counterparts without shear, to $\sim5\%$. The fact that these tests all result in similar values seems to indicate that the effect of shear dominates over the other types of perturbations. These tests have image distance ratios consistent with those of the observed quad population.

\subsubsection{Extended source fitting}
The results for the fitting using an array of sources are similar, with some exceptions. The ``1ell'' and ``2ell'' results are consistent with the point-source values. The ``OffCent'', ``Fourier'', and All results for $h$ have moved upward by 1.6 $\sigma$, 1.2 $\sigma$, and 2.4 $\sigma$ respectively, now consistent with an unbiased value of $h$. While the inclusion of extended source information has reduced the bias in these cases, there is still quite large scatter ($\sim10-20\%$). 

The $\gamma$ tests return values which are consistent with the point-source fittings, still biased $\sim10\%$ upward, although they are no longer all within 1 $\sigma$ of each other. All 9 tests result in a median value which is biased upward relative to the ``1ell'' result.

\vspace{15pt}

Curious if the level of bias is related to the radial offset (in the tests where the centers are offset), we searched for a correlation between $h$ and $r_{offset}$ and found no correlation for any of the relevant tests ($R< 0.1$).  Large offset radius tends to reduce the fraction of good fits and slightly increase the scatter of $h$, but the relationship between offset radius and $h$ is not  straightforward to predict across many lenses. 

% \begin{figure}
%  \centering
%  \includegraphics[width=\linewidth]{hvsoffsetradius.png}
%     \caption{{\color{cyan}figure shows lack of correlation between offset center and h. Not sure where to put within text} 
%           }
% \label{fig:hvsoffr}
% \end{figure}

\subsection{Ramifications for $H_0$} \label{ssec:hramifications}

The tests in this paper recover a wide range of values for $h$. Here we will parse these results and determine what lessons can carry forward to real measurements of $h$.

First, when we focus on the tests without shear, we note that all four perturbations to a simple elliptical shape increase the scatter considerably over the ``1ell'' case. 
When the only complexity is a misaligned position angle between the light and dark matter distributions, the result is still consistent with an unbiased value  for the point-source fitting, but the other three types of perturbations can cause significant bias in addition to the scatter  in both fittings. 
The tests with offset centers (``OffCent'' and ``All'') recover the most biased values of $h$.  We note that the direction of the bias is not the same for these tests across the two fittings. This result, in conjunction with the large scatter of these tests, implies that the recovery of $h$ for lenses with offset centers can depend considerably on the fitting method, with no guarantee of an accurate or precise value.

% {\color{cyan} (old text, remove) This is likely related to the known effect of profile slope on $h$ through the MSD \citep{Gomer19}. The offset centers spread out the mass distribution, causing an effectively shallower slope, which returns a lower value of $h$. This result shows the effects of mass contour shape and profile slope are quite intertwined.}

Meanwhile, unlike the tests without shear, the main result of the tests with shear is that both fittings return more or less the same value of $h$, independent of the other additional perturbations to the elliptical shape. Lenses which produce quads consistent with the observed radial image ratios (i.e. those with shear) result in $h$ being biased upward by $\sim10\%$. In fact, at least for the extended source fitting, all perturbations with or without shear result in this upward bias. A concerning implication of this result is that the observed population of quads may also recover $h$ biased upward by a similar amount. It is worth noting that the H0LiCOW value of $H_0$ at present is 8.9\% higher than the Planck value \citep{HC13,Planck18}, meaning that a bias of this order could explain the discrepancy. Because the quads which best match real systems are biased by an amount similar to the observed $h$ discrepancy, it is possible that the puzzles of the observed image distance ratios and $h$ may be related. The role of shear in this puzzle will be discussed in Sections \ref{ssec:roleofshear} and \ref{ssec:radratiocauses}.

It is interesting to consider the effect of our $\chi^2/dof<1$ selection. Because we assumed quite optimistic error bars on image position and time delay measurements for the point-source fitting, our requirement of $\chi^2/dof<1$ is likely more stringent than any existing survey. If surveys of real systems made a similar selection, they would probably allow more of the whole sample into their good sample due to larger uncertainties and therefore more acceptable fits. As such, it is likely that the resulting distribution of $h$ would be somewhere between our good sample and our whole sample results. In general, the scatter increases considerably in the whole sample, especially in cases with offset centers. The more drastic change in this case happens because a smaller fraction of systems for these tests is fit with $\chi^2/dof<1$. For the tests with shear, the good sample is biased higher in $h$ than the whole sample, but with significantly less scatter.
This selection is less relevant for the extended-source fitting for several reasons, chiefly that since a higher fraction of quads are fit with $\chi^2/dof<1$, there is less difference between the good sample and the whole sample.

\subsection{Role of shear}\label{ssec:roleofshear}
Clearly the role of external shear is important in the context of lens models. The tests we have done here include values of shear as high as 0.4 (uniformly distributed between 0 and 0.4), which is almost certainly non-astrophysical if one interprets shear as representative of external mass. Shears  even more extreme than this are necessary to explain the relative radial positions of quad images, but we make no claims that this much mass exists outside the lens systems. Instead, we argue that shear is a stand-in parameter for the inadequacies of the ellipse model, adding in a type of perturbation to it. We have tried to introduce every physically-motivated form of perturbation to this model that we can think of (for a fixed 1D density profile), but none of them have been able to reproduce the observed radial image ratios as well as shear can. The puzzle of how to reproduce these ratios will continue beyond this paper, although we will speculate as to potential causes in Section \ref{ssec:radratiocauses}. For now, we will not dwell on the large values that shear can take here and instead consider it just another parameter to fit.

For all tests with shear, the correlation between recovered shear and input shear is very good. In all cases, it is as strong or stronger than the correlation between input and recovered ellipticity-- when external shear is present, it is recovered well. This likely happens because the model explicitly includes shear, so the model is looking for the right kind of perturbation to an ellipse. Compare this to ellipticity, where we measure the correlation between the baryon ellipticity and the ellipticity of the total recovered mass model, which is a different, albeit related, quantity. The correlation in the recovery of ellipticity is weaker than that of shear for all tests.

An interesting result is that for the point-source fitting the tests with shear have more quads with good $\chi^2/dof$ and also better-recovered ellipticities than those without shear. All measures of goodness of fit are improved, even though the lenses themselves are actually more complicated than those without shear. It appears that the fitting procedure is better tuned for finding shear and is better able to handle other perturbations when shear is present. Perhaps the presence of quads which have images at different radii has allowed for a better fit. Additionally, for both fittings, the scatter in $h$ has decreased for all tests (except the ``1ell$\gamma$'' for the point-source fitting), so the presence of shear has largely made the result more consistent.

Motivated to make a comparison to observables, we return to our model-free statistical measures of a population of quads: the radial position of images relative to the outermost image and the deviation from the FSQ resulting from the angular distribution of images. \citet{WW15} showed that deviations from the FSQ for a single system can be caused by external shear or by deviations to the elliptical shape \citep{Gomer18}, while we have shown that shear can also reproduce the distribution or radial ratios. If shear were responsible for both quantities, then the innermost radial ratio and the deviation from the FSQ should be correlated. Meanwhile a lack of correlation would imply that the two measures are affected by different physical means. This correlation would be observable without the need for any fitting processes, using simply the radial and angular image positions. We will explore the utility of this test for a population of systems.

We measured this correlation for synthetic quads from our tests and for the observed population. In particular, it is useful to compare ``1ell$\gamma$'', where deviations from the FSQ are caused solely by shear, ``All'', where deviations from the FSQ are caused by perturbations to the elliptical shape instead of shear, and ``All$\gamma$'', where both forms of deviation are present. Quads from ``1ell$\gamma$'' return a moderate correlation with a Pearson R of -0.47 (a small radial ratio correlates with a large $\Delta \theta_{23}$), while the ``All'', and ``All$\gamma$'' tests result in no correlation. The fact that ``All$\gamma$'' results in no correlation unfortunately means that this test cannot diagnose the presence of shear-- ``All$\gamma$'' had shear but returned no correlation. Rather, this test returns a correlation if shear is the only cause of deviations from the FSQ. When we measure the observed quad population, we recover no correlation. We conclude that shear is not the sole cause of deviations from the FSQ, confirming the results of \citet{WW15}and \citet{Gomer18}, but this test is inconclusive regarding the degree to which shear contributes to deviations from the FSQ. In this manner, the ``All'' tests are again the most similar to the observed population of quads, but this test cannot distinguish between``All'' or ``All$\gamma$''.

\subsection{Possible causes of extreme radial image ratios}\label{ssec:radratiocauses}
Considering that none of the perturbations to the elliptical shape have had as substantial an effect on the radial image ratios as shear, we can only speculate as to a few other candidates, as well as discuss why they may be unlikely.

To begin, we can explore some of the most extreme cases and how they have been modeled in previous work. The most extreme ratio of our sample of quads is the B1422+231 system. Attempts to model this system with one ellipse have been insufficient. \citet{Hogg94} used 2 SIS halos and included external galaxies as point masses to get a $\chi^2/dof$ of 16. \citet{Kundic97} fit the system with an SIS+shear model to get a $\chi^2$ of 40.3 with a shear of 0.23. \citet{Raychaudhury03} estimated the external shear contribution of the nearby group to be between 0.16 and 0.67. Clearly this system is more complicated than a field elliptical galaxy. Another extreme system is SDSS J002240, which has been fit by \citet{Allam07} with a very large ellipticity of 0.53, significantly larger than the values we consider in this paper. RXJ 0911+0551 has a nearby cluster which provides a minimum external shear of 0.15 \citep{Burud98}. Meanwhile B2045+265 has been modeled extensively as it has very anomalous flux ratios. \citet{McKean07} modeled the system as 2 SIEs+shear, with a $\chi^2/dof$ of 1.9 and a shear of 0.2. The authors suspect substructure, and additionally note that the lens potential is likely more boxy than the elliptical mass model, since the shear does not correspond to the nearby galaxy. More recently, \citet{Spingola18} modeled the same system as part of the SHARP program and found the ellipticity misaligned with the light and that the shear changes direction when group galaxies are included, suspecting additional complexity in the mass distribution beyond parametric models. To sum up, of the systems with the most extreme radial ratios, many attempts to model them frequently result in extreme ellipticities or shears, or other complications to the shape, or poor $\chi^2$ fits.

% \citet{Raychaudhury} either small gamma, small alpha, or large gamma, large alpha (alpha is power law slope)

Many of these extreme cases have nearby groups of clusters. It seems logical that the go-to candidate for these extreme image ratios would be the most commonly accepted physical interpretation of shear: the effect of mass external to the system. External mass certainly contributes to shear, but may not provide enough shear to solve this mystery. To match the observed population, we used synthetic quads with up to 0.4 shear, which still doesn't quite reach the most extreme ratios in Figure \ref{fig:imageratios}, but can statistically match the population of systems. We can estimate size of a group necessary to produce this shear by considering the external shear of a singular isothermal sphere (SIS) with lensing potential
\begin{equation}
   \psi(\vec{\theta}) = \frac{D_{ls}}{D_s} \frac{4\pi\sigma^2}{c^2} \vec{\theta}
\end{equation}
and taking derivatives to evaluate the shear
\begin{flalign*}
    \gamma_1&=\frac{1}{2}\left(\frac{\partial^2\psi}{\partial\theta_x^2}-\frac{\partial^2\psi}{\partial\theta_y^2}\right) ,\\
    \gamma_2&=\frac{\partial^2\psi}{\partial\theta_x\partial\theta_y} ,
\end{flalign*} 
\begin{equation}\label{eq:SISshear}
    \gamma=\sqrt{\gamma_1^2+\gamma_2^2}=\frac{D_{ls}}{D_s}\frac{2\pi\sigma^2}{c^2\vec{\theta}} 
\end{equation}
Using our source and lens redshifts (3.0 and 0.6, respectively) and converting units, we find that $\gamma=0.376\left(\frac{\sigma}{200\text{ km s}^{-1}}\right)^2\theta^{-1}$, where $\sigma$ is in km/s and $\theta$ is in arcseconds. To produce an external shear of 0.4 from 1 arcsecond away, a SIS would require a velocity dispersion of just over 200 km/s. From 1 arcminute away, an unrealistic dispersion of 1600 km/s is required. As an example, one of the observed systems with large external shear is PG 1115+080, which has a group about 10 arcseconds away with a velocity dispersion of 390 km/s \citep{Wilson16}. Using Equation \ref{eq:SISshear} for the redshifts of PG 1115+080, the group likely contributes an external shear of 0.15, roughly matching most attempts to model the system \citep{Keeton97,Treu02,Chen19}. It is difficult to imagine a system with 2-3 times more shear than PG 1115+080, as it would need a similar-sized group to either be 2-3 times closer to the lens or a group with a velocity dispersion nearly twice as large. This rather extreme case seems close to the upper limit on the extent to which external structure can contribute to shear.

% It may also be possible that line-of-sight structure at a different redshift could affect image positions differently, requiring less mass to create a similar shear.

One factor to consider is that the amount of shear necessary to produce extreme radial ratios depends somewhat on profile slope. Lenses with shallower slopes have less-concentrated mass and require less external shear to deflect images and produce extreme ratios. A rough estimate of the magnitude of this effect can be evaluated by considering several fits to the same system by \citet{HC12}. Table C1 lists the fit values of slope and shear for seven variants of the H0LiCOW power law ellipse+shear model for the WFI2033-4723 system. The values of slope and shear are correlated, with 3D slope ranging from 1.90 to 2.02 and shear ranging from 0.109 to 0.126. Since these fits are on the same system, any possible effects of quad configuration are controlled, and we can roughly conclude that a slope change of 0.1 produces a shear change of approximately 0.015. The model we use for these tests has a 3D slope (before perturbations are added) of approximately 2.14 \citep{Gomer19}. If real lens systems had shallower slopes, they would not require  a shear of 0.4 or greater to match the observed quads. If, for example, the slope were 1.9, a shear of up to 0.36 would be required, assuming the effect from  WFI2033-4723 scales similarly to these extreme cases, an assumption which merits caution. Even so, a shear of 0.36 is still too high to be plausibly caused by external mass.

One possibility is that some fraction of lenses are not elliptical galaxies, but actually edge-on disks. \citet{Hsueh16,Hsueh17} have shown for two particular systems (B1555+375 and B0712+472, respectively) that edge-on disks are able to explain flux ratio anomalies. It is possible that a disk could masquerade as a high ellipticity or shear, but seems unlikely that a disk could have sufficient mass to drastically alter image positions. The two cases from \citet{Hsueh16,Hsueh17} require the disk to constitute $\simeq15\%$ of the mass within the Einstein radius. These systems have innermost radial image ratios of 0.82 and 0.61, respectively, still leaving the most extreme cases with innermost ratios $< 0.4$ unexplained. In addition, this would only affect the fraction of lenses which happen the be spirals and also happen to be edge-on. Turning to simulations, \citet{Hsueh18} found that edge-on disks can introduce astrometric anomalies of 3 mas in 13\% of lenses in the Illustris simulation. 

Because this is an effect of spiral galaxies, it would be more common in systems with smaller masses and therefore Einstein radii. Based on Figure 2 of \citet{Hsueh18}, a selection of lenses with Einstein radii greater than 1 arcsecond would select mostly elliptical galaxies. A comparison of the distributions of radial image ratios between galaxies with Einstein radii larger or smaller than 1 arcsecond could help illuminate the degree to which edge-on disks play a role. When comparing these subsets of the observed sample of 50 systems (listed in Table \ref{table:quadlist}), the lenses with smaller image radii, hypothesized to be more commonly spirals, tend to actually have less extreme image ratios (mean 0.71) than those systems with larger radii (mean 0.61). A notable exception is B1422+231 (discussed above), which happens to be the system with the most extreme innermost ratio of 0.23 and a mean image radius of only 0.84 arcseconds. A KS Test of these two distributions returns a p-value of 5.9\%, so  these two subsets are consistent with having been drawn from the same population. If there is any trend, the trend goes the wrong way, with spirals having less extreme ratios, meaning it is unlikely that  edge-on disks are responsible for extreme radial image ratios.

The most extreme ratios come from only a few systems, so it is possible that a combination of effects could happen for these systems. Perhaps a system with a high (but still plausible) ellipticity and similarly high (but still plausible) shear like 0.15 in the same direction as the ellipticity axis, on a profile with a shallower than isothermal slope could produce image ratios which are as extreme as the tail in Figure \ref{fig:imageratios}. Or perhaps these systems can be ruled out as outliers due to nearby groups, mergers, edge-on disks, or other significant additions to the ellipse+shear model, although the systems with $r_i/r_{max}\lesssim0.8$ still require explanation. A selection bias which preferentially selects high-ellipticity systems likely plays a role. A metastudy is merited which fully explores how each of these systems are fit and the ways in which the population of systems can or cannot be explained in the context of the ellipse+shear model, which is beyond the scope of this paper.

\section{Conclusion}
When lens systems are fit with ellipse+shear models, it is implicitly assumed that the mass distribution is elliptical. However, there is statistical evidence that the observed quad population comes from lens mass distributions which are more complicated than a simple elliptical shape. Inclusion of $\Lambda$CDM substructure, even if all clump masses are increased by a factor of 10, does not resolve the issue. A mismatch between the true mass distribution and the model used to fit it can alter the recovery of parameters such as $H_0$. 

We created a series of tests in which lenses with perturbations to the elliptical shape are fit with an ellipse+shear model and compared the recovered values of $H_0$.  Following the prescription of \citet{Gomer18}, we produced a synthetic quad population which matches the statistical properties of the observed azimuthal distribution of quad images. We simultaneously attempted to create a population which matches the statistical properties of the observed radial image positions, with mixed success. When fitting these populations, biases on $H_0$ of order 10\% or more can result, depending on the type of asymmetry being considered. Kinematic constraints are not included. The distributions of $H_0$ values are shown in Figure \ref{fig:hwhisker}.  The most significant perturbation in terms of influencing recovered $H_0$ is the mass dipole with respect to the center of light. This is also the perturbation needed to reproduce the statistical distribution of relative polar image angles of observed quads. More generally, this illustrates a danger of parametric models when accuracy is required at the percent level: parametric models use assumptions about mass distributions to combat degeneracies, but can return incorrect results if the assumptions are incorrect.

To carry out a fair comparison between observed and mock quads, one needs to make sure that the statistical properties of the image distribution around the lens center are the same for both samples. As such, we discuss some interesting statistical properties of the observed population of quads and make comparisons to our synthetic test populations. The most critical quantity of interest is the ratio of image distances relative to the farthest-out image. This measure of radial spread of the images can be quite extreme for the observed quad population-- more extreme than can be reproduced with astrophysically reasonable values of external shear. This property of the observed population of quads has not been discussed in the literature and is difficult to explain. It adds to the evidence that real systems have more complicated mass distributions than ellipse+shear. We speculate as to some possible causes of these extreme image ratios, but a fully-realized reproduction of the statistical properties of the quad population is a task for future work.

\section*{Acknowledgments}

This project has received funding from the European Research Council (ERC) under the European Union’s Horizon 2020 research and innovation programme (COSMICLENS : grant agreement No 787886).

\section*{Data Availability Statement}
The data underlying this article will be shared on reasonable request to the corresponding author.

\appendix
\section{Table of observed quads used}
\begin{table*}
\centering
\setlength\tabcolsep{5pt}
\begin{tabular}{c c }
%\multicolumn{2}{c}{List of observed systems} \\
%\hline
System & Reference(s) \\
\hline
MG 2016+112   &  \citet{Lawrence84}; \citet{Nair97};  \citet{Koopmans02}; \\
B 0712+472    &  CLASS \citet{Fassnacht02}; \citet{Jackson98}\\
B 2045+265   &  CLASS \citet{Fassnacht99}; \citet{McKean07}; \citet{Sluse12}\\
B 1933+503 lobe &  \citet{Nair98}\\
SLACS J2300+002 & \citet{Ferreras08}; figure 6.60 in \citet{Bolton08}\\
MG 0414+0534  & CASTLES \citet{Falco99}\\
SLACS J1636+470 &  \citet{Ferreras08}; figure 6.58 in \citet{Bolton08}\\
HS 0810+2554  & CASTLES \citet{Falco99} \\
B 1555+375 &   \citet{Marlow99}; \citet{Barvainis02}\\
PG 1115+080  &   \citet{Miranda07} \\
J 100140.12+020  040.9 &  \citet{Jackson08}\\
SDSS J1330+1810 &  \citet{Oguri08}\\
SLACS J1205+491 &  \citet{Ferreras08}; figure 6.38 in \citet{Bolton08}\\
B 1422+231  &  CASTLES \citet{Falco99} \\
WFI 2026-4536  & CASTLES \citet{Falco99} \\
CLASS B1359+154 &   \citet{Myers99}; \citet{Rusin00})\\
RXJ 0911+0551 & \citet{Burud98} \\
SDSS J1538+5817 &  \citet{Grillo10}\\
SDSS J125107  &   \citet{Kayo07} \\
RXJ 1131-1231  & \citet{Morgan06})\\
SDSS J120602.09 & \citet{Lin09}\\
WFI 2033-4723  & CASTLES \citet{Falco99} \\
SDSS J002240  & \citet{Allam07}; \citet{Dessauges-Zavadsky11}\\
J 095930.94+023  427.7 &  \citet{Jackson08}\\
HE 0230-2130  & \citet{Wisotzki99}\\
SDSS 1402+6321  &  \citet{Bolton05}\\
SDSS 0924+0219 &   \citet{Keeton06}\\
LSD Q0047-2808 &  \citet{Koopmans03}; \citet{Brewer06}\\
B 1933+503 core  &  \citet{Nair98}\\
B 1608+656  & CASTLES \citet{Falco99} (center G1 has filter dependent position) \\
SDSS 1138+0314 & CASTLES \citet{Falco99} \\
Q 2237+0305  & CASTLES \citet{Falco99} \\
HE 1113-0641  &  \citet{Blackburne08}\\
HST 14113+5211  & \citet{Lubin00}\\
H 1413+117   &  \citet{MacLeod09}\\
HST 14176+5226 & CASTLES \citet{Falco99} \\
HST 12531-2914  & CASTLES \citet{Falco99} \\
HE 0435-1223    &  \citet{Kochanek06}; \citet{Courbin11}\\
SDSS 1011+0143  & CASTLES \citet{Falco99}\\
SLACS J0946+006 &  \citet{Gavazzi08}; \citet{Vegetti10}\\
GRAL113100-441959 &  \citet{Krone-Martins18}\\
GRAL203802-400815  &  \citet{Krone-Martins18}\\
GRAL122629-454209 &  \citet{Krone-Martins18}\\
J1606-2333 & \citet{Lemon18}\\
J1721+8842 & \citet{Lemon18}\\
ATLAS 0259-1635 & \citet{Schechter18} (galaxy position is from modeling)\\
B0128+437  & \citet{Lagattuta10}\\
KiDS0239-3211 & \citet{Sergeyev18}\\
DES J0408-5354  & \citet{Agnello17}; \citet{Shajib20}\\
DES J0405-3308 & \citet{Anguita18}\\

\end{tabular}
\caption{
    A list of the systems used when referring to the observed quad population. Systems which have a well-defined lens center (such that image angles and distances can be calculated) were selected from a variety of surveys. Figure \ref{fig:FSQcompare} uses the only quads from \citet{WW12}, which are the first 40 quads in this table, omitting the bottom 10.}
\label{table:quadlist}
\end{table*}
%%%%%%%%%%%%%%%%%%%%%%%%%%%%%%%%%%%%%%%%%%%%%%%%%%

% \paragraph{Note added.} This is also a good position for notes added
% after the paper has been written.

% The bibliography will probably be heavily edited during typesetting.
% We'll parse it and, using the arxiv number or the journal data, will
% query inspire, trying to verify the data (this will probalby spot
% eventual typos) and retrive the document DOI and eventual errata.
% We however suggest to always provide author, title and journal data:
% in short all the informations that clearly identify a document.

% \begin{thebibliography}{99}

% \bibitem{a}
% Author, \emph{Title}, \emph{J. Abbrev.} {\bf vol} (year) pg.

% \bibitem{b}
% Author, \emph{Title},
% arxiv:1234.5678.

% \bibitem{c}
% Author, \emph{Title},
% Publisher (year).

% Please avoid comments such as "For a review'', "For some examples",
% "and references therein" or move them in the text. In general,
% please leave only references in the bibliography and move all
% accessory text in footnotes.

% Also, please have only one work for each \bibitem.

% \end{thebibliography}

\bibliographystyle{mnras}
\bibliography{ellpaperdraft}
% Don't change these lines
\bsp	% typesetting comment
\label{lastpage}
\end{document}